%% file: main.tex
\documentclass[aps,prl,twocolumn,amsmath,amssymb,floatfix,longbibliography]{revtex4-2}
	
\usepackage{color}
\usepackage{mathrsfs}
\usepackage{graphicx}
\usepackage{dcolumn}
\usepackage{bm}
\usepackage{hyperref}
\usepackage{amsmath,amsfonts,amssymb,ulem}
\usepackage{epstopdf}
\usepackage{xcolor}

\usepackage{graphicx}

\begin{document}

\title{Retrieving transient magnetic fields of ultrarelativistic laser plasma via ejected \\electron polarization}

\author{Zheng Gong}
\email[]{gong@mpi-hd.mpg.de}
 \affiliation{Max-Planck-Institut f\"{u}r Kernphysik, Saupfercheckweg 1, 69117 Heidelberg, Germany}
\author{Karen Z. Hatsagortsyan}
\email[]{k.hatsagortsyan@mpi-hd.mpg.de}
 \affiliation{Max-Planck-Institut f\"{u}r Kernphysik, Saupfercheckweg 1, 69117 Heidelberg, Germany}
\author{Christoph H. Keitel}
 \affiliation{Max-Planck-Institut f\"{u}r Kernphysik, Saupfercheckweg 1, 69117 Heidelberg, Germany}

\date{\today}

\begin{abstract}

Interaction of an ultrastrong short laser pulse with non-prepolarized near-critical density plasma is investigated in an ultrarelativistic regime, with an emphasis on the radiative spin polarization of ejected electrons. Our particle-in-cell simulations show explicit correlations between the angle resolved electron polarization and the structure and properties of the transient quasistatic plasma magnetic field. While the magnitude of the spin signal is the indicator of the magnetic field strength created by the longitudinal electron current, the asymmetry of electron polarization is found to gauge the island-like magnetic distribution which emerges due to the transverse current induced by the laser wave front. Our studies demonstrate that the spin degree of freedom of ejected electrons could potentially serve as an efficient  tool to retrieve the features of strong plasma fields.

\end{abstract}

\maketitle
Magnetic fields play a crucial role in various plasma collective phenomena and nonlinear quantum electrodynamic processes in extreme environments of laboratory and universe~\cite{lai2001matter,gamma_ray_burst,extreme_environment}. The astrophysical magnetic fields can govern the internal structure of interstellar shocks~\cite{Magnetic_field_amp_in_shock_2021}, mediate the radio wave emission nearby neutron stars~\cite{origin_radio_emission}, induce baryon inhomogeneities~\cite{baryon_inhomogeneities}, and catalyse the dark matter formation~\cite{axion_Dark_matter_detection}. Self-generated fields with strength $\sim10^4$ Tesla have been produced in high-intensity plasma experiments~\cite{B_gen_nature2002_tatarakis,B_gen_PRL2012_sarri,B_gen_PRL2013_schumaker,turbulent_NP2014_meinecke,Weibel_NP2015_huntington}, and the guidance of jet flows by laboratory magnetic fields helps interpret the evolution of young stellar objects~\cite{albertazzi2014laboratory_sci,revet2017laboratory_sci_adv,revet2021laboratory_nc}. With recent advancement of ultrastrong laser techniques~\cite{Vulcan10,ELI,ElI-Beamlines,XCELS,danson2019petawatt,yoon2019achieving,yoon2021realization} more extreme conditions and larger fields are expected in ultrarelativistic laser plasma interaction~\cite{mourou2006optics,marklund2006nonlinear,bell2008possibility,di2012extremely,RRD_bulanov2015,poder2018experimental,cole2018experimental}.

Generally, detection of plasma magnetic fields requires an external probe beam, where the field information is imprinted in the velocity space of charged particles~\cite{detect_li2007observation,detect_willingale2010fast,detect_macchi2013ion,detect_lancia2014topology,detect_zhang2020measurements,proton_imaging_bott2021time} or the rotated polarization vector of the optical beam~\cite{FR_tatarakis2002measurements,FR_kaluza2010measurement,FR_buck2011real,FR_zhou2018self}. However, these conventional methods are inapplicable for scenarios with unprecedented field strength, ultrashort time scale ($\sim$fs), and overcritical plasma density~\cite{wang2019structured}. Furthermore, the spin, an intrinsic property of particles, offers a new degree of freedom of information, which is widely utilized in exploring magnetization of solids~\cite{jungwirth2014spin}, nucleon structure~\cite{pol_e_nuclear_structure_1}, and phenomena beyond the standard model~\cite{pol_e_beyound_stand_model}. In extreme laser fields there is a strong coupling of the electron spin to the laser magnetic field \cite{Walser_2002,Sorbo_2017,Seipt_2018,sorbo2018ppcf,li2020production}, which may yield radiative spin polarization (SP)~\cite{Sokolov_1968,Baier_1967,Baier_1972,Derbenev_1973}, i.e., polarization of  electrons  due to spin-flip during photon emissions. Even though in the oscillating field the electron net SP is suppressed, fast polarization of a lepton beam with laser pulses becomes possible when the symmetry of the monochromatic field is broken, such as in an elliptically polarized, or in two-color laser pulses \cite{li2019ultrarelativistic_PRL,li2019electron_PRApp,Wan_2019plb,chen2019polarized,Seipt_2019}. Because of collective effects, more complex spin dynamics occurs in strong laser field interaction with plasma.
Consequently, the question arises if it is possible to employ the spin signal of spontaneously ejected particles from plasma to retrieve information on transient plasma fields.

\begin{figure}
\includegraphics[width=0.5\textwidth]{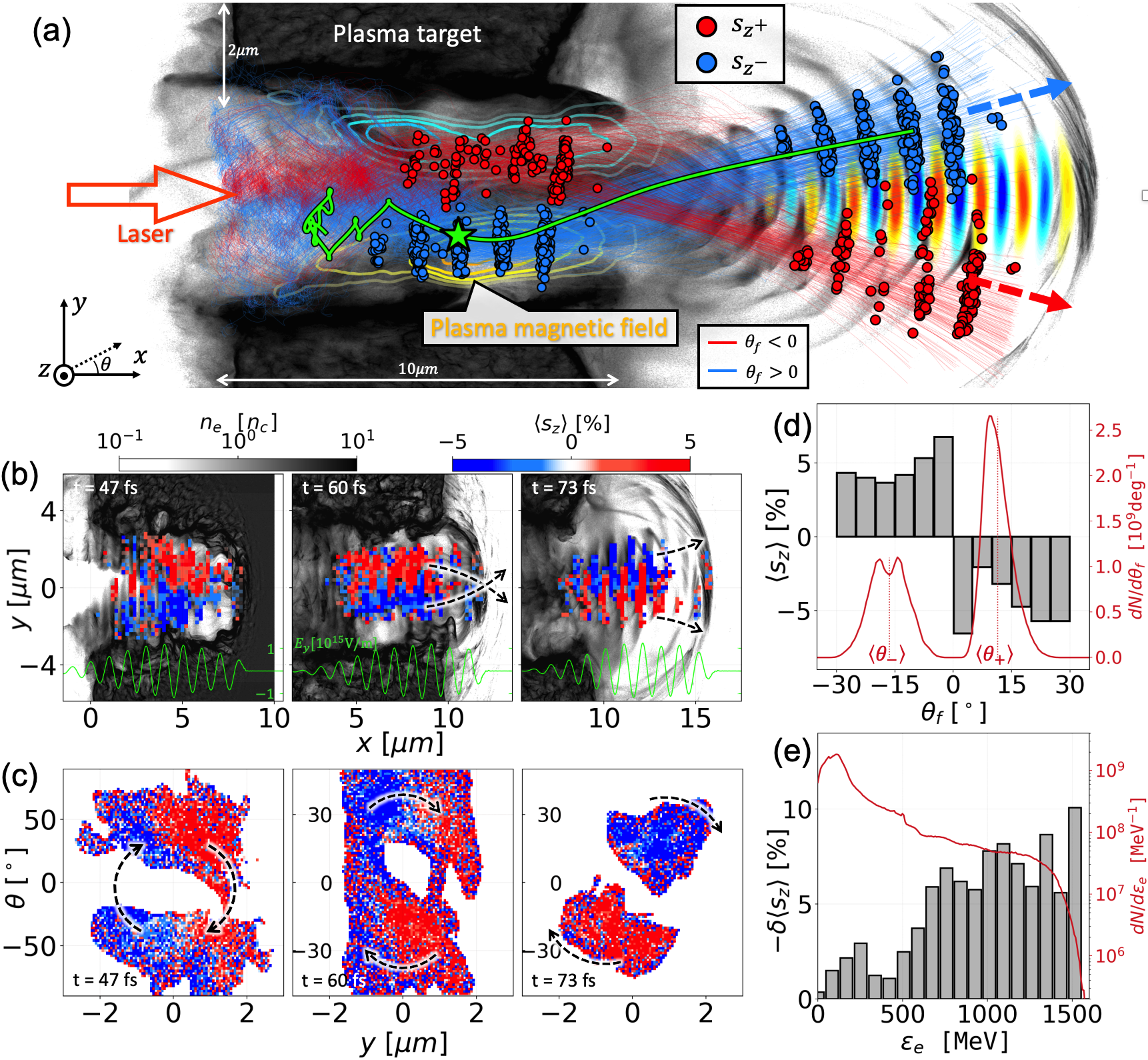}
\caption{(a) The interaction scheme: the laser pulse impinges on unpolarized plasma (the electron density is shown in  gray shades);  accelerated and radiatively polarized electrons due to spin-flips form outgoing polarized bunches. The red (blue) dots represent the electrons with spin $s_z$\,$=$\,$1$ ($-1$) and the lines show their movement tendency. The green line shows a typical electron trajectory with a spin-flip marked by a pentagram. (b) and (c) show snapshots of the electron SP distribution in spatial $(x,y)$ coordinates and transverse $(y,\theta)$ phase space, respectively. The green lines in (b) profile the laser field $E_y$ at slice $y=0$. (d) Angular distribution of electron number $dN/d\theta_f$ and SP $\left<s_z\right>$. (e) $\delta\left<s_z\right>$ and  $dN/d\varepsilon_e$ vs electron energy $\varepsilon_e$. All parameters are indicated in the text.}
\label{fig1}
\end{figure}

\begin{figure}
\includegraphics[width=1\columnwidth]{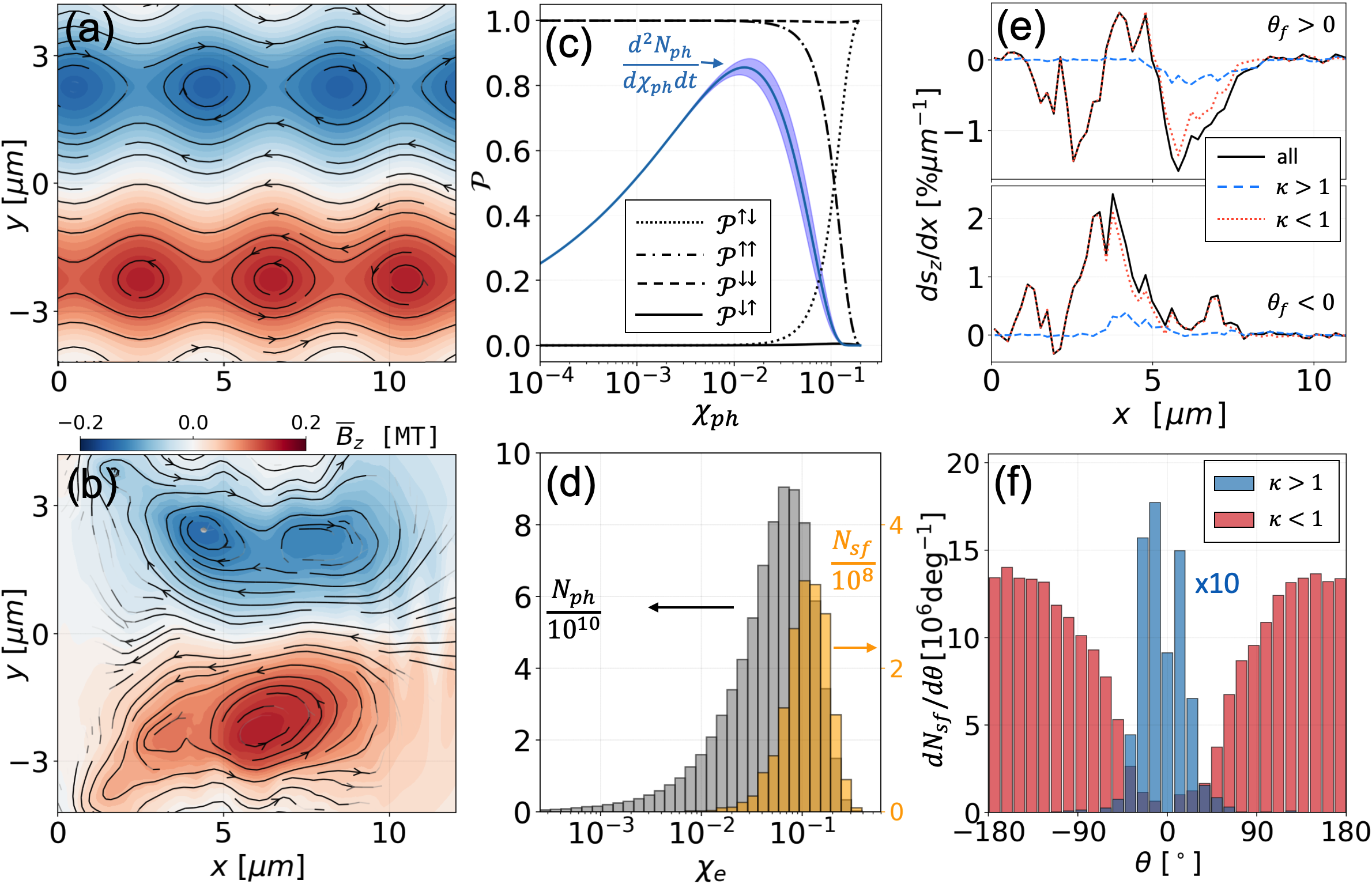}
\caption{The $\overline{B}_z$ obtained from (a) analytical theory and (b) PIC simulations, where the black arrows denote the direction of electric current $\bm{j}$. (c) Probabilities $\mathcal{P}$ of electron spin-flip after emitting a photon with $\chi_{ph}$. The $\uparrow\downarrow$ represents the spin-flip from parallel to antiparallel with respect to the magnetic field direction. The blue line profiles the photon emission probability $d^2N_{ph}/d\chi_{ph} dt$ with a bandwidth accounting for the influence of electron spin. (d) The number distribution of all emitted photons $N_{ph}$ (grey) and emission associated with spin-flips $N_{sf}$ (yellow). (e) The spatial dependent SP differentiate $ds_z/dx$ contributed by the cases of $\kappa\lessgtr1$ for electrons with final angle $\theta_f>0$ and $\theta_f<0$, respectively. (f) The angular dependence of spin-flip occurrence, where the result of condition $\kappa>1$ is multiplied by 10 for better visibility.}
\label{fig2}
\end{figure}

In this Letter, based on particle-in-cell (PIC) simulations, we investigate the ultrarelativistic dynamics of a short strong pulse interacting with a non-prepolarized near-critical density plasma, see Fig.~\ref{fig1}. Special attention is devoted to describing the spin dynamics of plasma electrons, being strongly disturbed by the radiative spin-flips modulated by the quasistatic plasma magnetic field (QPMF). The latter is commonly transient with a time scale as short as the driving pulse duration while being quasistatic with respect to the fast oscillating laser field. We show that the angle dependent SP of ejected electrons  carries signatures of the inhomogeneous QPMF. The signal of SP of ejected electrons can be used to predict the strength of the leading order antisymmetic QPMF created by the longitudinal current. A more detailed analysis reveals that the asymmetry of SP of two outgoing divergent electron bunches characterizes the secondary QPMF, which is induced by a transverse transient current and generally neglected in previous studies~\cite{pukhov1996relativistic,pukhov1999_DLA,Stark2016_PRL,gong2020direct,hussein2021towards}. The sum of these two part QPMFs gives rise to a nonlinear island-like magnetic structure [see Fig.~\ref{fig2}(a)(b)].  Our results demonstrate that the spin degree of freedom of ejected electrons from ultrarelativistic plasmas can be employed in principle as a tool to retrieve information on the QPMF structure and properties.

In 2D PIC simulations, a near-critical density target is irradiated by a linearly polarized pulse (with the transverse electric field along $y$). Our main example adopts a peak intensity of $1.7 \times 10^{23}\mathrm{W/cm}^2$, equivalent to the normalized field amplitude $a_0= 350$ given the laser wavelength $\lambda_0=1\mu m$. The pulse has a $2.6\,\mu m$ focal spot size and $18\,$fs duration (FWHM intensity measure). The target has thickness $l_0=10\,\mu m$ and electron (carbon) density $n_e=5n_c$ ($n_i=n_e/6$), where $n_{c}= m_e \omega_0^2 / 4 \pi e^2$ is the plasma critical density for a laser frequency $\omega_0=ck_0$; $m_e$ ($e$)  the electron mass (charge); $c$  the speed of light. The dynamics of spin precession is governed by the Thomas-Bargmann-Michel-Telegdi equation and spin-dependent photon emissions have been implemented in the EPOCH code~\cite{arber2015contemporary}, see ~\cite{SM}.

When the pulse impinges on the target, a fraction of bulk electrons is expelled outwards by the laser ponderomotive force to form a plasma channel~\cite{pukhov2002strong}. Meanwhile, the peripheral electrons are prone to be injected~\cite{ji2014_PRL} and subsequently polarized inside the channel due to spin-flips during photon emissions, see Fig.~\ref{fig1}(b). Since the ion reaction partially compensates the transverse charge separation~\cite{jansen2018leveraging}, the quasistatic electric field $\overline{E}_y$ is negligible in this scenario. Thus the deflection of the accelerated electrons in transverse direction is predominantly governed by the azimuthal QPMF $\overline{B}_z$, which is presumably sustained by the longitudinally forward moving electron current $j_x$. The simulation results in Fig.~\ref{fig1}(a) show that the electrons with a positive (negative) final angle $\theta_f$ mainly originate from the plasma region of $y<0$ ($y>0$). As the magnetic field $\overline{B}_z\sim -\mu_0 |j_0|y$ is antisymmetric, created by the nearly uniform current $j_x\sim -|j_0|$, the electrons exiting the plasma area with a final angle $\theta_f>0$ mostly experience a positive $\overline{B}_z$ [see Fig.~\ref{fig1}(a)] and vice versa. This leads to oppositely SP ejected electron bunches: $\left<s_z\right><0$ ($\left<s_z\right>>0$) for the electron bunch of $\theta_f>0$ ($\theta_f<0$). The spatial evolution of  SP in Fig.~\ref{fig1}(b) manifests that two groups of electrons are firstly polarized and confined inside the channel, and then intersect with each other towards the opposite transverse direction. This procedure is also unveiled by the evolution of SP  $\left<s_z\right>$ in the transverse phase space $(y,\theta)$ in Fig.~\ref{fig1}(c), where $\theta=\arctan(p_y/p_x)$ denotes the direction of electron momentum. The clockwise rotation of $\left<s_z\right>$ indicates that the QPMF not only generates spatial dependent SP but also deflects the electrons to form an angle dependent polarization distribution of ejected electrons. In Fig.~\ref{fig1}(d), asymmetry exists for both electron SP and number angular distributions. Specifically, the averaged SP (final angle) with a positive $\theta_f$ is $\left<s_+\right>\approx-3.3\%$ ($\left<\theta_+\right>\approx11.4^\circ $), whereas $\left<s_-\right>\approx4.0\%$ ($\left<\theta_-\right>\approx-16.5^\circ $) for $\theta_f<0$. The magnitude of the SP signal is characterized by the parameter $\delta\left<s_z\right>\equiv\left<s_+\right>-\left<s_-\right>$. According to Fig.~\ref{fig1}(e), SP is insignificant for low-energy electrons because of damped radiative spin-flips. Therefore, the criterion of $\varepsilon_e>4a_0m_ec^2$ is adopted here to filter out the low-energy noise. To reveal more subtle features of  QPMF $\overline{B}_z$, we introduce also the spin (angle) asymmetry characteristics via the absolute difference: $\Delta \left<s_z\right> \equiv |\left<s_+\right>| - |\left<s_-\right>|$ and $\Delta \left<\theta\right> \equiv |\left<\theta_+\right>| - |\left<\theta_-\right>|$, which will be discussed below.

The QPMF $\overline{B}_z$ is determined by electric currents via $\partial \overline{B}_z/\partial y=\mu_0 j_x$ and $\partial \overline{B}_z/\partial x=-\mu_0 j_y$ (with the vacuum permeability $\mu_0$). 
In general, inside a laser-driven plasma channel, the current is dominated by the longitudinal one $j_x$ and the transverse current $j_y$ is neglected~\cite{pukhov1996relativistic,pukhov1999_DLA,Stark2016_PRL,gong2020direct,hussein2021towards}. 
However, the magnetic field in our simulation shows an irregular structure, with multiple islands associated with the current kinks and vortices, see Fig~\ref{fig2}(b). The latter indicates that the transverse current $j_y$ is important in characterizing the exact form of $\overline{B}_z$. Let us divide QPMF into two parts $\overline{B}_z=\overline{B}_{z,1}+\overline{B}_{z,2}$, where the leading part $\overline{B}_{z,1}$ is induced by  $j_x$, while the secondary $\overline{B}_{z,2}$ by $j_y$: $\partial \overline{B}_{z,1}/\partial y=\mu_0 j_x$ and $\partial \overline{B}_{z,2}/\partial x=-\mu_0 j_y$. The leading part $\overline{B}_{z,1}\sim-\mu_0|e|n_ec y$ with antisymmetric feature $\overline{B}_{z,1}(-y)=-\overline{B}_{z,1}(y)$ is ubiquitously utilized in previous studies~\cite{pukhov1996relativistic,pukhov1999_DLA,Stark2016_PRL,gong2020direct,hussein2021towards}. Now, we focus on the secondary $\overline{B}_{z,2}$. Considering the electron velocity $v_y=p_y/(\gamma m_ec)$ and momentum $p_y\sim A_y=a_0\cos (\xi+\phi_0)$ where $\xi=\omega_0 t-k_0x$ and $\phi_0$ the carrier envelop phase (CEP), we obtain $j_y\approx-|e|\int n_2\delta (x/v_g-t) v_y dt\approx |e|n_2 \cos[\omega(x/v_g-x/v_{ph})+\phi_0]$. The $\delta (t-x/v_g)$ function indicates that the transverse current is predominantly contributed by the electron density $n_2\delta (t-x/v_g)$ piled up at the front edge of the plasma channel nearby the region $x\sim v_gt$, where the electron's transverse velocity is significant. Here, $v_g$ ($v_{ph}$) is the laser group (phase) velocity in plasma, and the Lorentz-factor $\gamma\sim a_0$ is assumed. 
With $\partial \overline{B}_{z,2}/\partial x = -\mu_0j_y$, the secondary magnetic field can be estimated:
\begin{eqnarray}\label{eq:B_2}
\overline{B}_{z,2}\sim -\frac{\mu_0|e|n_2}{k_2}\sin(k_2x+\phi_0),
\end{eqnarray}
where $k_2=k_0(v_{ph}-v_g)/v_g$. The analytically predicted $\overline{B}_z=\overline{B}_{z,1}+\overline{B}_{z,2}$ is shown in Fig.~\ref{fig2}(a), which agrees qualitatively with the simulated $\overline{B}_z$ in Fig.~\ref{fig2}(b). The asymmetric periodic island-like structure of QPMF $\overline{B}_z$ stems from the nontrivial current vortex $(\nabla \times \bm{j})_{x,y}\neq0$ generated by the transverse current of electrons plough away by the laser beam front.

\begin{figure}[t]
\includegraphics[width=1\columnwidth]{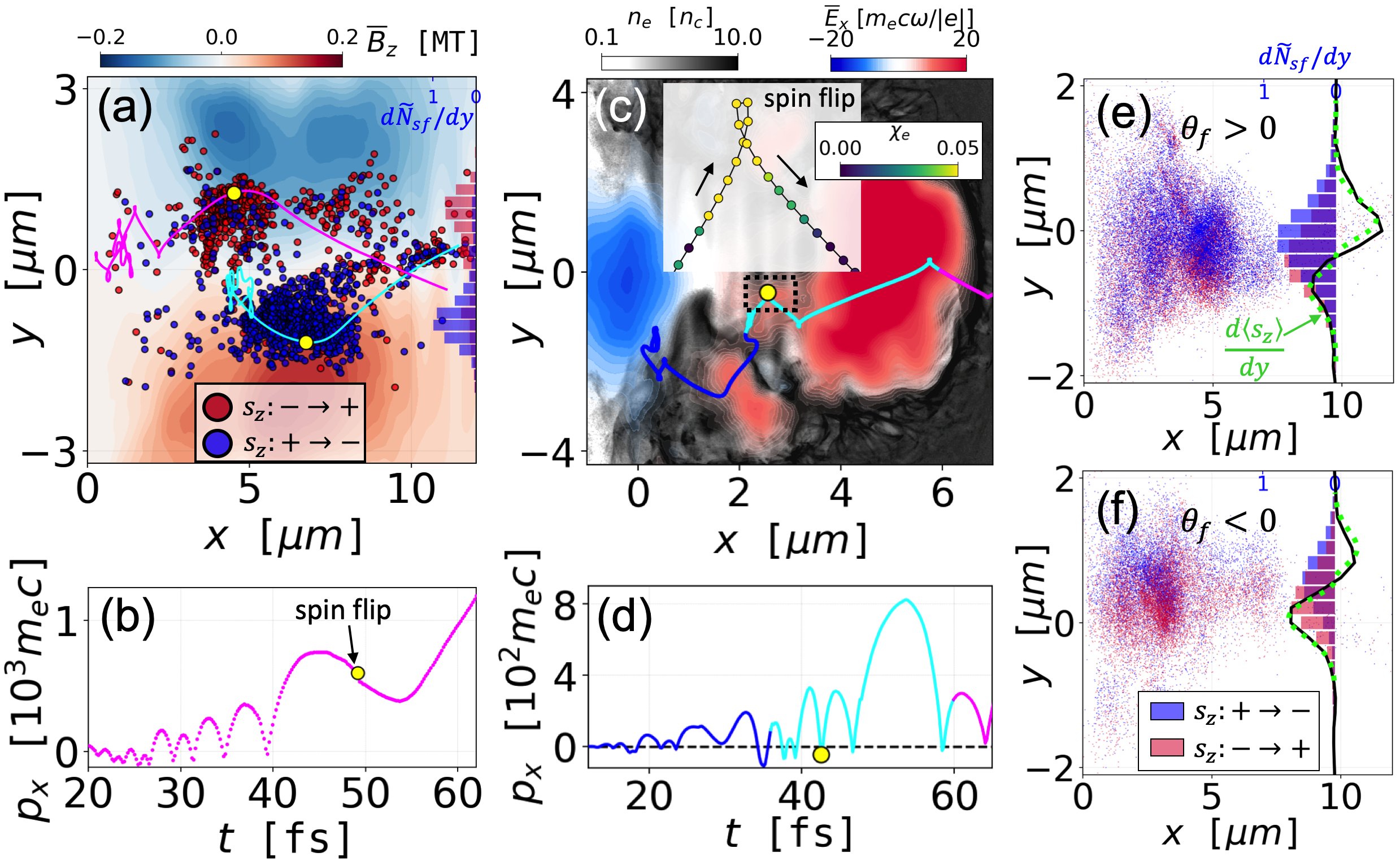}
\caption{(a) The red (blue) dots present the electron spin-flip from $s_z=-1$ ($1$) to $1$ ($-1$) in the plasma field dominant regime ($\kappa>1$) and the histogram exhibits its dependence on the transverse coordinate $d\Tilde{N}_{sf}/dy$. The magenta line refers to a typical electron trajectory and its time evolution of $p_x$ is in (b). (c) The typical electron trajectory for the regime $\kappa<1$ and its corresponding momentum evolution in (d). The spin-flips of electron trajectories in (a)-(d) are marked by yellow circles. (e) and (f) show the spin-flips with $\kappa<1$ for electrons with final angle $\theta_f>0$ and $\theta_f<0$, where the solid black (dashed lime) lines profile the simulated (analytically derived) dependence of net electron SP $s_z$ on the coordinate $y$.}
\label{fig3}
\end{figure}

As we are interested in the relation of the electron SP to the magnetic field structure, and considering the polarization attributable to the spin-flip during a photon emission, we analyze the probability of this process $\mathcal{P}(\chi_{ph})$ in Fig.~\ref{fig2}(c) for typical parameters of our PIC simulations. Here, the electron with an initial $\gamma_e=2000$ normally crosses the uniform magnetic field $B_0=10^4$T, and the electron quantum invariant parameter is $\chi_e\sim 0.1$, with $\chi_{e,ph}\equiv (e\hbar/m_e^3c^4) |F_{\mu\nu} p^{\nu}|$ and the momentum $p^{\nu}$ of the electron or photon, respectively. As Fig.~\ref{fig2}(c) illustrates, the electron spin-flips exclusively take place when emitting an energetic photon with $\chi_{ph}$ close to $\chi_e$, while the photon emission probability is peaked at $\hbar\omega_c\sim \chi_e\gamma_em_ec^2$ (at $\chi_e< 1$), i.e., the peak of the spin-flip process is shifted with respect to the photon emission to higher $\chi_e$'s, see Fig.~\ref{fig2}(d). Both the laser magnetic field $B_l$ and QPMF $\overline{B}_z$ can cause the electron spin-flip as $\chi_e\approx\gamma_e |[(1-\cos\theta)B_l+\overline{B}_z]|/B_c$ with the Schwinger limit $B_c\approx4\times 10^9\,$T. We introduce the parameter $\kappa=|\overline{B}_z/[(1-\cos\theta)B_l|]$, defining two regimes, when the electron spin-flip is dominated by the plasma ($\kappa>1$) or by the laser field  ($\kappa<1$). The evolution of SP in Fig.~\ref{fig2}(e) demonstrates that the laser field dominated regime ($\kappa<1$)  mostly contributes to the final electron SP.
A distinguishable feature between the $\kappa\lessgtr 1$ regimes is the angle $\theta$ of the electron's instantaneous momentum when the spin-flip occurs. As the angular dependent spin-flip shows in Fig.~\ref{fig2}(f), the $\kappa<1$ regime applies at backward  emissions, while $\kappa>1$ for forward ones.

\begin{figure}[t]
\includegraphics[width=1\columnwidth]{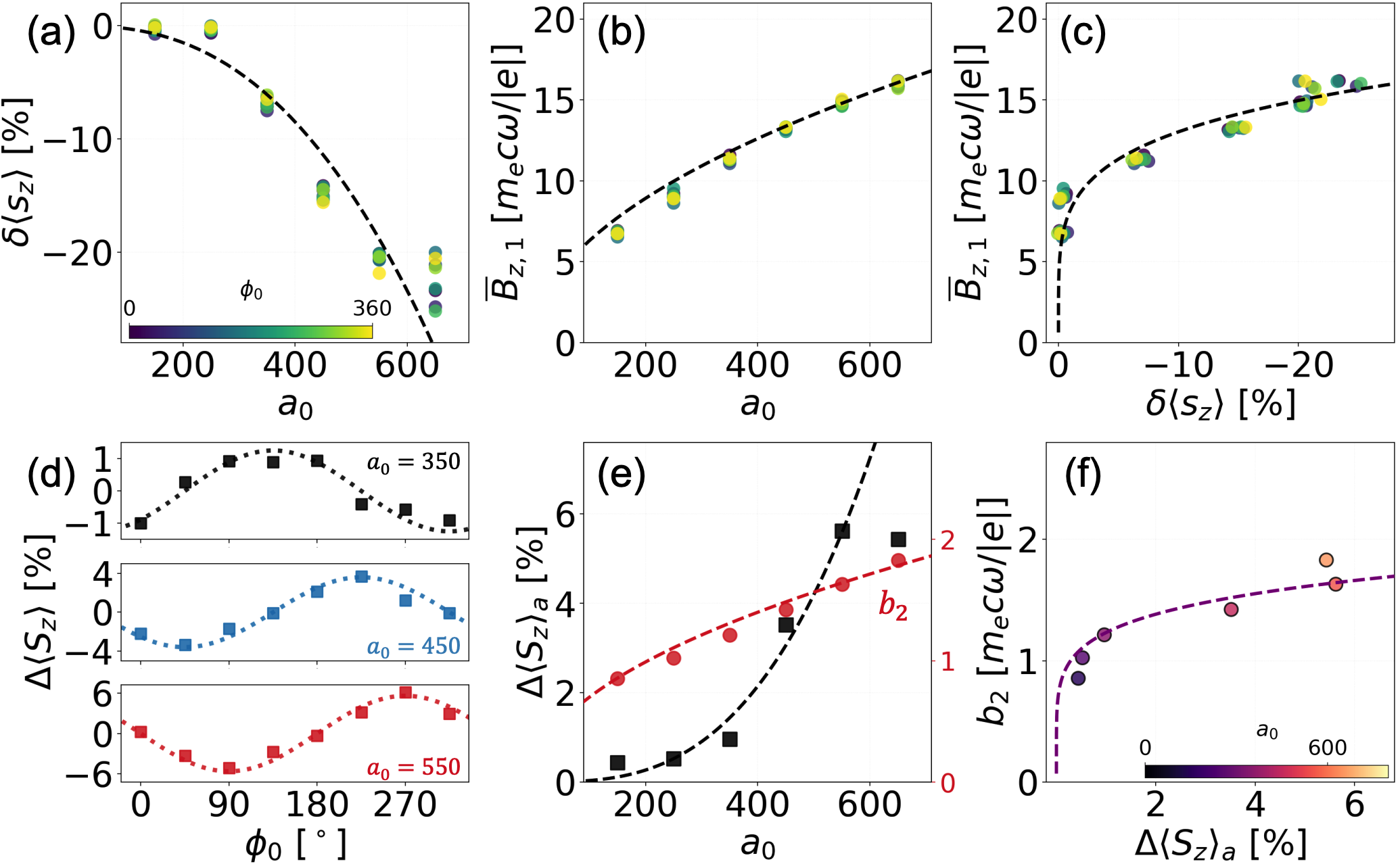}
\caption{The dependence of (a) electron SP $\delta\left<s_z\right>$ and (b) the leading order QPMF $\overline{B}_{z,1}$ on $a_0$. (c) The correlation between $\delta \left<s_z\right>$ and $\overline{B}_{z,1}$. In (a)-(c), the dots refer to the simulation results with different laser CEP $\phi_0$ while the dashed black line denotes the theory. (d) The dependence of $\Delta \left<s_z\right>$ on $\phi_0$ for different $a_0$. (e) The dependence of $\Delta \left<s_z\right>_a$ and $b_2$ on $a_0$. (f) The correlation between the  amplitude  of  the  SP  asymmetry $\Delta \left<s_z\right>_a$ and the secondary QPMF $b_2$.}
\label{fig4}
\end{figure}
The detailed particle tracking  further confirms these conclusions. In the $\kappa>1$ regime Figs.~\ref{fig3}(a),(b), the position of spin-flip with $\kappa>1$ is closely correlated with the spatial distribution of QPMF $\overline{B}_z$. The time evolution of  $p_x$ illustrates that the spin-flip happens after the electron starts an efficient acceleration and its velocity aligns longitudinally $\theta\ll 1$, resulting in $(1-\cos\theta)B_l<\overline{B}_z$. For the laser dominant regime $\kappa<1$, the electron trajectory and momentum evolution [in Fig.~\ref{fig3}(c)(d)] demonstrate that the typical spin-flip occurs at the electron's temporarily backward motion, when  $(1-\cos\theta)\sim 1$ and $B_l>\overline{B}_z$.

It should be noted that even in the laser dominant regime, the QPMF $\overline{B}_z$ is still the key factor for the SP. The reason is that the laser field has oscillating character. Although it can cause spin-flips, its net contribution to the final SP is negligible. The laser magnetic field $B_l$ acts as a catalyst to enhance the electron spin-flips by increasing $\chi_e$ and net spin-flips contributing to the final SP are still determined by $\overline{B}_z$~\cite{SM}. We may estimate $s_z\approx -\int\overline{B}_z/|B_l|\mathcal{A}(\chi_e)dt$ (at $B_l\gg \overline{B}_z) $, with $\mathcal{A}(\chi_e)=(\sqrt{3}\alpha_f m_ec^2\chi_e)/(h\gamma_e)\mathcal{A}^*(\chi_e)$,  and $\mathcal{A}^*(\chi_e)\approx 0.18\chi_e$ (at $0.01<\chi_e<0.4$)~\cite{SM}. The electrons with final angle $\theta_f>0$ ($\theta_f<0$) are mainly exposed to the QPMF $\overline{B}_z>0$ ($\overline{B}_z<0$) at the region $y<0$ ($y>0$), and the overall SP  with $\theta_f>0$ ($\theta_f<0$) would be $s_z<0$ ($s_z>0$) which are illustrated as the solid black lines in Fig.~\ref{fig3}(e)(f).

Thus, we calculate the electron's SP magnitude $\delta \left<s_z\right>$ being correlated with the leading order QPMF $\overline{B}_{z,1}$:
\begin{eqnarray}\label{eq:sz_a}
\delta \left<s_z\right>\sim -\eta \frac{|e||\overline{B}_{z,1}|}{m_e\omega_0}a_0^2,
\end{eqnarray}
where $\gamma_e\sim a_0$ is used, and $\eta\approx 4\times10^{-8}$ accounts for the deviations from the radiative spin evolution. With $\overline{B}_{z,1}\approx \sqrt{(a_0/4\pi) (n_e/n_c)}(m_e\omega_0/|e|)$, we find the SP scaling  $\delta \left<s_z\right>\propto a_0^{5/2}$, as well as the relation $\overline{B}_{z,1}\approx [-(\delta \left<s_z\right>/\eta)(n_e/4\pi n_c)^2]^{1/5}$. In Fig.~\ref{fig4}(a)(b),(c), the analytically predicted scalings of $\delta \left<s_z\right>$ and $\overline{B}_{z,1}$ are in good accordance with the 2D simulation results.

Finally, we show how with the help of the SP asymmetry signal $\Delta\left<s_z\right>$ defined above, the secondary QPMF can be retrieved. In the $\Delta\left<s_z\right>$ signal the contribution of the $\overline{B}_{z,1}$ is cancelled, and $\Delta \left<s_z\right> \approx 2\int (\overline{B}_{z,2}/|B_l|)\mathcal{A}(\chi_e) dt$. Since $\overline{B}_{z,2}\sim b_2\sin(k_2x+\phi_0)$ is oscillating along the longitudinal position (along the laser CEP), the overall effect of $\overline{B}_{z,2}$ imprinted on the signal of $\Delta \left<s_z\right>$ is oscillating as well. Taking into account the results for $\delta \left<s_z\right>$ and $\overline{B}_{z,2}$, we find for the asymmetry signal
\begin{eqnarray}\label{eq:D_sz_a}
\Delta \left<s_z\right>\sim \delta \left<s_z\right> \frac{b_2}{|\overline{B}_{z,1}|}\frac{k_0}{k_2} \cos(k_2l_0+\phi_0),
\end{eqnarray}
where $b_2$ is the amplitude of $\overline{B}_{z,2}$. The oscillating dependence of $\Delta \left<s_z\right>$ on the laser CEP $\phi_0$ is reproduced by the simulation results in Fig.~\ref{fig4}(d). We see that the amplitude of the SP asymmetry signal $\Delta \left<s_z\right>_a$ is directly related to the secondary QPMF $b_2$: $\Delta \left<s_z\right>_a\sim |\delta \left<s_z\right>|(b_2/|\overline{B}_{z,1}|)(k_0/k_2)\sim \eta\sqrt{n_c/n_e}b_2a_0^{5/2}$ [see Fig.~\ref{fig4}(e)], where $k_0/k_2$$\sim$$(a_0n_c/n_e)^{1/2}$ is obtained from simulation results~\cite{SM}. Moreover, the amplitude of the secondary quasistatic magnetic field $b_2\approx 0.03\sqrt{a_0(n_e/n_c)}(m_e\omega_0/|e|)$ can be estimated through the number conservation between the initial undisturbed plasma and the electrons piled up at the front of the channel edge. Then, the correlation between  $\Delta \left<s_z\right>_{a}$ and $b_2$ is established: $b_2 \sim 0.12[\Delta \left<s_z\right>_{a}/\eta]^{1/6}$, which is in reasonable agreement with the simulation results [see Fig.~\ref{fig4}(f)]. Therefore, the SP signals of $\delta \left<s_z \right>$ and $\Delta \left<s_z\right>$ allow to retrieve the strength of the leading and secondary QPMFs.

\begin{figure}
\includegraphics[width=0.98\columnwidth]{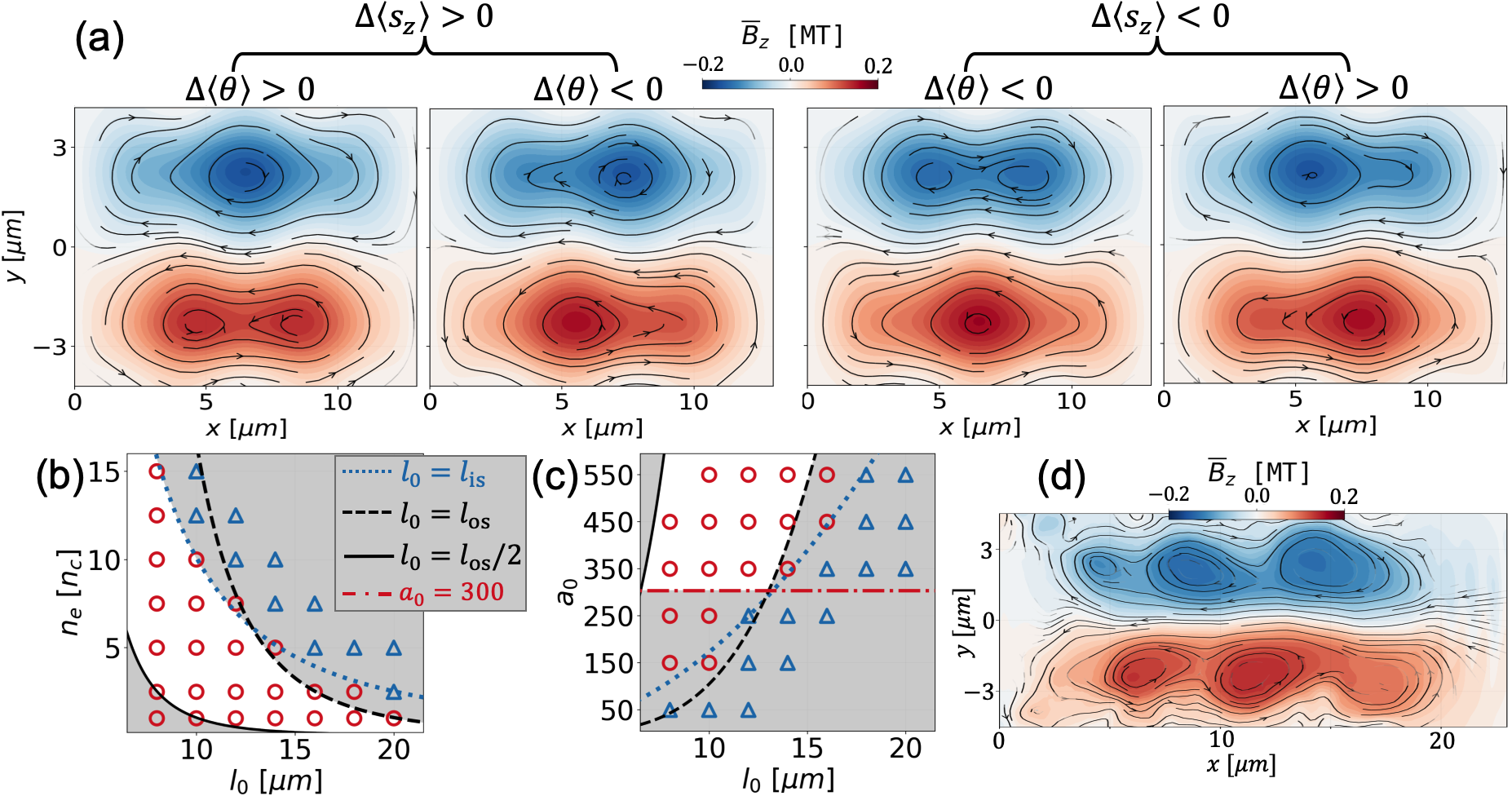}
\caption{(a) The $\overline{B}_z$ predicted by the model based on the sign of $\Delta \left<s_z\right>$ and $\Delta \left<\theta\right>$. The valid range of the model is illustrated in the white region of the parameter space in (b) $a_0=350$ and (c) $n_e=5n_c$, where the circles (triangles) mark the simulation results of $\overline{B}_z$ with no more (more) than two islands on each side of $y=0$. (d) The $\overline{B}_z$ obtained from simulation for the case of $a_0=350$, $n_e=5n_c$, and $l_0=20\mu m$. }
\label{fig5}
\end{figure}

In addition, the combination of  $\Delta \left<s_z\right>$ and $\Delta \left<\theta\right>$, allows to predict the concrete spatial structure of $\overline{B}_z$, see Fig.~\ref{fig5}(a). Based on the sign of $\Delta \left<s_z\right>$ and $\Delta \left<\theta\right>$, the analytically estimated magnetic island structures agree well with the simulation results, see \cite{SM}. 
We define the limitations of the presented field retrieval model. Firstly, it is applicable when no more than two QPMF islands exist at $y\lessgtr0$, with a criterion $l_0<l_\mathrm{island}\sim 1.7\lambda_0(a_0n_c/n_e)^{1/2}$~\cite{SM}. Secondly, to exclude the influence of depolarization, the ejected electrons should experience a half-period of betatron oscillation inside the channel, with a criterion $0.5l_\mathrm{os}<l_0<l_\mathrm{os}\sim 5\lambda_0(a_0n_c/n_e)^{1/4}$~\cite{SM}. Consequently, the valid range of the model is $0.5l_\mathrm{os}<l_0<\mathrm{min}\{l_\mathrm{is},\,l_\mathrm{os}\}$ shown as the white area in Fig.~\ref{fig5}(b)(c).
Our method based on the electron radiative polarization will be efficient in the quantum radiation dominated regime at $\alpha_f a_0\chi_e\gtrsim 1$ (approximately at $a_0\gtrsim 300$) and $\chi_e\gtrsim 0.1$~\cite{di2012extremely}, with a SP signal within the precision of the electron polarimetry of $\sim 0.4\%$~\cite{narayan2016precision}. The requirement might be relieved at alternative setups~\cite{cole2018experimental}, e.g.,  in multiple colliding laser pulses~\cite{zhang2020relativistic}, where new schemes for the magnetic field retrieval may be needed.

To confirm the robustness of our scheme, we also investigated the role of experimental imperfections and uncertainties, in particular, the asymmetry in the driving laser pulse, and the ramp-up and -down of the plasma density profile~\cite{SM}. The simulation results indicate that the presented scheme is robust to moderate imperfections of such practical issues. It should be noted that distinguishing more complex field structures, e.g., the three-island structure like that in Fig.~\ref{fig5}(d), could be achievable with modifications of the retrieval method, see an example in \cite{SM}, which however needs further exploration.

In conclusion, the ejected electron spin provides a new degree of freedom to extract information on the structure and magnitude of different components of the transient plasma fields. Our results open a new avenue for the electron spin-based plasma diagnostics in extreme conditions, which are prevalent in astrophysical environments and are expected in near future laser facilities.


The original version of code EPOCH adapted here is funded by the UK EPSRC grants EP/G054950/1, EP/G056803/1, EP/G055165/1 and EP/ M022463/1. Z. Gong would like to thank A. V. Arefiev, X.-Q. Yan and J.-X. Li for useful discussions. The Supplemental Material includes Refs.~\cite{thomas1927kinematics,bargmann1959precession,duclous2010monte,elkina2011qed,ridgers2014modelling,gonoskov2015extended,ma2007directly,bin2015ion,wang2011laser}.


\input{output.bbl}
\end{document}

%% file: output.bbl
%

%% file: main.bbl
\begin{thebibliography}{78}%
\makeatletter
\providecommand \@ifxundefined [1]{%
 \@ifx{#1\undefined}
}%
\providecommand \@ifnum [1]{%
 \ifnum #1\expandafter \@firstoftwo
 \else \expandafter \@secondoftwo
 \fi
}%
\providecommand \@ifx [1]{%
 \ifx #1\expandafter \@firstoftwo
 \else \expandafter \@secondoftwo
 \fi
}%
\providecommand \natexlab [1]{#1}%
\providecommand \enquote  [1]{``#1''}%
\providecommand \bibnamefont  [1]{#1}%
\providecommand \bibfnamefont [1]{#1}%
\providecommand \citenamefont [1]{#1}%
\providecommand \href@noop [0]{\@secondoftwo}%
\providecommand \href [0]{\begingroup \@sanitize@url \@href}%
\providecommand \@href[1]{\@@startlink{#1}\@@href}%
\providecommand \@@href[1]{\endgroup#1\@@endlink}%
\providecommand \@sanitize@url [0]{\catcode `\\12\catcode `\$12\catcode
  `\&12\catcode `\#12\catcode `\^12\catcode `\_12\catcode `\%12\relax}%
\providecommand \@@startlink[1]{}%
\providecommand \@@endlink[0]{}%
\providecommand \url  [0]{\begingroup\@sanitize@url \@url }%
\providecommand \@url [1]{\endgroup\@href {#1}{\urlprefix }}%
\providecommand \urlprefix  [0]{URL }%
\providecommand \Eprint [0]{\href }%
\providecommand \doibase [0]{https://doi.org/}%
\providecommand \selectlanguage [0]{\@gobble}%
\providecommand \bibinfo  [0]{\@secondoftwo}%
\providecommand \bibfield  [0]{\@secondoftwo}%
\providecommand \translation [1]{[#1]}%
\providecommand \BibitemOpen [0]{}%
\providecommand \bibitemStop [0]{}%
\providecommand \bibitemNoStop [0]{.\EOS\space}%
\providecommand \EOS [0]{\spacefactor3000\relax}%
\providecommand \BibitemShut  [1]{\csname bibitem#1\endcsname}%
\let\auto@bib@innerbib\@empty
\bibitem [{\citenamefont {Lai}(2001)}]{lai2001matter}%
  \BibitemOpen
  \bibfield  {author} {\bibinfo {author} {\bibfnamefont {D.}~\bibnamefont
  {Lai}},\ }\bibfield  {title} {\bibinfo {title} {Matter in strong magnetic
  fields},\ }\href@noop {} {\bibfield  {journal} {\bibinfo  {journal} {Reviews
  of Modern Physics}\ }\textbf {\bibinfo {volume} {73}},\ \bibinfo {pages}
  {629} (\bibinfo {year} {2001})}\BibitemShut {NoStop}%
\bibitem [{\citenamefont {Piran}(2005)}]{gamma_ray_burst}%
  \BibitemOpen
  \bibfield  {author} {\bibinfo {author} {\bibfnamefont {T.}~\bibnamefont
  {Piran}},\ }\bibfield  {title} {\bibinfo {title} {The physics of gamma-ray
  bursts},\ }\href@noop {} {\bibfield  {journal} {\bibinfo  {journal} {Reviews
  of Modern Physics}\ }\textbf {\bibinfo {volume} {76}},\ \bibinfo {pages}
  {1143} (\bibinfo {year} {2005})}\BibitemShut {NoStop}%
\bibitem [{\citenamefont {Ruffini}\ \emph {et~al.}(2010)\citenamefont
  {Ruffini}, \citenamefont {Vereshchagin},\ and\ \citenamefont
  {Xue}}]{extreme_environment}%
  \BibitemOpen
  \bibfield  {author} {\bibinfo {author} {\bibfnamefont {R.}~\bibnamefont
  {Ruffini}}, \bibinfo {author} {\bibfnamefont {G.}~\bibnamefont
  {Vereshchagin}},\ and\ \bibinfo {author} {\bibfnamefont {S.-S.}\ \bibnamefont
  {Xue}},\ }\bibfield  {title} {\bibinfo {title} {Electron--positron pairs in
  physics and astrophysics: from heavy nuclei to black holes},\ }\href@noop {}
  {\bibfield  {journal} {\bibinfo  {journal} {Physics Reports}\ }\textbf
  {\bibinfo {volume} {487}},\ \bibinfo {pages} {1} (\bibinfo {year}
  {2010})}\BibitemShut {NoStop}%
\bibitem [{\citenamefont {Bohdan}\ \emph {et~al.}(2021)\citenamefont {Bohdan},
  \citenamefont {Pohl}, \citenamefont {Niemiec}, \citenamefont {Morris},
  \citenamefont {Matsumoto}, \citenamefont {Amano}, \citenamefont {Hoshino},\
  and\ \citenamefont {Sulaiman}}]{Magnetic_field_amp_in_shock_2021}%
  \BibitemOpen
  \bibfield  {author} {\bibinfo {author} {\bibfnamefont {A.}~\bibnamefont
  {Bohdan}}, \bibinfo {author} {\bibfnamefont {M.}~\bibnamefont {Pohl}},
  \bibinfo {author} {\bibfnamefont {J.}~\bibnamefont {Niemiec}}, \bibinfo
  {author} {\bibfnamefont {P.~J.}\ \bibnamefont {Morris}}, \bibinfo {author}
  {\bibfnamefont {Y.}~\bibnamefont {Matsumoto}}, \bibinfo {author}
  {\bibfnamefont {T.}~\bibnamefont {Amano}}, \bibinfo {author} {\bibfnamefont
  {M.}~\bibnamefont {Hoshino}},\ and\ \bibinfo {author} {\bibfnamefont
  {A.}~\bibnamefont {Sulaiman}},\ }\bibfield  {title} {\bibinfo {title}
  {Magnetic field amplification by the weibel instability at planetary and
  astrophysical shocks with high mach number},\ }\href@noop {} {\bibfield
  {journal} {\bibinfo  {journal} {Phys. Rev. Lett.}\ }\textbf {\bibinfo
  {volume} {126}},\ \bibinfo {pages} {095101} (\bibinfo {year}
  {2021})}\BibitemShut {NoStop}%
\bibitem [{\citenamefont {Philippov}\ \emph {et~al.}(2020)\citenamefont
  {Philippov}, \citenamefont {Timokhin},\ and\ \citenamefont
  {Spitkovsky}}]{origin_radio_emission}%
  \BibitemOpen
  \bibfield  {author} {\bibinfo {author} {\bibfnamefont {A.}~\bibnamefont
  {Philippov}}, \bibinfo {author} {\bibfnamefont {A.}~\bibnamefont
  {Timokhin}},\ and\ \bibinfo {author} {\bibfnamefont {A.}~\bibnamefont
  {Spitkovsky}},\ }\bibfield  {title} {\bibinfo {title} {Origin of pulsar radio
  emission},\ }\href@noop {} {\bibfield  {journal} {\bibinfo  {journal} {Phys.
  Rev. Lett.}\ }\textbf {\bibinfo {volume} {124}},\ \bibinfo {pages} {245101}
  (\bibinfo {year} {2020})}\BibitemShut {NoStop}%
\bibitem [{\citenamefont {Jedamzik}\ and\ \citenamefont
  {Pogosian}(2020)}]{baryon_inhomogeneities}%
  \BibitemOpen
  \bibfield  {author} {\bibinfo {author} {\bibfnamefont {K.}~\bibnamefont
  {Jedamzik}}\ and\ \bibinfo {author} {\bibfnamefont {L.}~\bibnamefont
  {Pogosian}},\ }\bibfield  {title} {\bibinfo {title} {Relieving the hubble
  tension with primordial magnetic fields},\ }\href@noop {} {\bibfield
  {journal} {\bibinfo  {journal} {Phys. Rev. Lett.}\ }\textbf {\bibinfo
  {volume} {125}},\ \bibinfo {pages} {181302} (\bibinfo {year}
  {2020})}\BibitemShut {NoStop}%
\bibitem [{\citenamefont {Foster}\ \emph {et~al.}(2020)\citenamefont {Foster},
  \citenamefont {Kahn}, \citenamefont {Macias}, \citenamefont {Sun},
  \citenamefont {Eatough}, \citenamefont {Kondratiev}, \citenamefont {Peters},
  \citenamefont {Weniger},\ and\ \citenamefont
  {Safdi}}]{axion_Dark_matter_detection}%
  \BibitemOpen
  \bibfield  {author} {\bibinfo {author} {\bibfnamefont {J.~W.}\ \bibnamefont
  {Foster}}, \bibinfo {author} {\bibfnamefont {Y.}~\bibnamefont {Kahn}},
  \bibinfo {author} {\bibfnamefont {O.}~\bibnamefont {Macias}}, \bibinfo
  {author} {\bibfnamefont {Z.}~\bibnamefont {Sun}}, \bibinfo {author}
  {\bibfnamefont {R.~P.}\ \bibnamefont {Eatough}}, \bibinfo {author}
  {\bibfnamefont {V.~I.}\ \bibnamefont {Kondratiev}}, \bibinfo {author}
  {\bibfnamefont {W.~M.}\ \bibnamefont {Peters}}, \bibinfo {author}
  {\bibfnamefont {C.}~\bibnamefont {Weniger}},\ and\ \bibinfo {author}
  {\bibfnamefont {B.~R.}\ \bibnamefont {Safdi}},\ }\bibfield  {title} {\bibinfo
  {title} {Green bank and effelsberg radio telescope searches for axion dark
  matter conversion in neutron star magnetospheres},\ }\href@noop {} {\bibfield
   {journal} {\bibinfo  {journal} {Phys. Rev. Lett.}\ }\textbf {\bibinfo
  {volume} {125}},\ \bibinfo {pages} {171301} (\bibinfo {year}
  {2020})}\BibitemShut {NoStop}%
\bibitem [{\citenamefont {Tatarakis}\ \emph
  {et~al.}(2002{\natexlab{a}})\citenamefont {Tatarakis}, \citenamefont {Watts},
  \citenamefont {Beg}, \citenamefont {Clark}, \citenamefont {Dangor},
  \citenamefont {Gopal}, \citenamefont {Haines}, \citenamefont {Norreys},
  \citenamefont {Wagner}, \citenamefont {Wei} \emph
  {et~al.}}]{B_gen_nature2002_tatarakis}%
  \BibitemOpen
  \bibfield  {author} {\bibinfo {author} {\bibfnamefont {M.}~\bibnamefont
  {Tatarakis}}, \bibinfo {author} {\bibfnamefont {I.}~\bibnamefont {Watts}},
  \bibinfo {author} {\bibfnamefont {F.}~\bibnamefont {Beg}}, \bibinfo {author}
  {\bibfnamefont {E.}~\bibnamefont {Clark}}, \bibinfo {author} {\bibfnamefont
  {A.}~\bibnamefont {Dangor}}, \bibinfo {author} {\bibfnamefont
  {A.}~\bibnamefont {Gopal}}, \bibinfo {author} {\bibfnamefont
  {M.}~\bibnamefont {Haines}}, \bibinfo {author} {\bibfnamefont
  {P.}~\bibnamefont {Norreys}}, \bibinfo {author} {\bibfnamefont
  {U.}~\bibnamefont {Wagner}}, \bibinfo {author} {\bibfnamefont {M.-S.}\
  \bibnamefont {Wei}}, \emph {et~al.},\ }\bibfield  {title} {\bibinfo {title}
  {Measuring huge magnetic fields},\ }\href@noop {} {\bibfield  {journal}
  {\bibinfo  {journal} {Nature}\ }\textbf {\bibinfo {volume} {415}},\ \bibinfo
  {pages} {280} (\bibinfo {year} {2002}{\natexlab{a}})}\BibitemShut {NoStop}%
\bibitem [{\citenamefont {Sarri}\ \emph {et~al.}(2012)\citenamefont {Sarri},
  \citenamefont {Macchi}, \citenamefont {Cecchetti}, \citenamefont {Kar},
  \citenamefont {Liseykina}, \citenamefont {Yang}, \citenamefont {Dieckmann},
  \citenamefont {Fuchs}, \citenamefont {Galimberti}, \citenamefont {Gizzi},
  \citenamefont {Jung}, \citenamefont {Kourakis}, \citenamefont {Osterholz},
  \citenamefont {Pegoraro}, \citenamefont {Robinson}, \citenamefont
  {Romagnani}, \citenamefont {Willi},\ and\ \citenamefont
  {Borghesi}}]{B_gen_PRL2012_sarri}%
  \BibitemOpen
  \bibfield  {author} {\bibinfo {author} {\bibfnamefont {G.}~\bibnamefont
  {Sarri}}, \bibinfo {author} {\bibfnamefont {A.}~\bibnamefont {Macchi}},
  \bibinfo {author} {\bibfnamefont {C.~A.}\ \bibnamefont {Cecchetti}}, \bibinfo
  {author} {\bibfnamefont {S.}~\bibnamefont {Kar}}, \bibinfo {author}
  {\bibfnamefont {T.~V.}\ \bibnamefont {Liseykina}}, \bibinfo {author}
  {\bibfnamefont {X.~H.}\ \bibnamefont {Yang}}, \bibinfo {author}
  {\bibfnamefont {M.~E.}\ \bibnamefont {Dieckmann}}, \bibinfo {author}
  {\bibfnamefont {J.}~\bibnamefont {Fuchs}}, \bibinfo {author} {\bibfnamefont
  {M.}~\bibnamefont {Galimberti}}, \bibinfo {author} {\bibfnamefont {L.~A.}\
  \bibnamefont {Gizzi}}, \bibinfo {author} {\bibfnamefont {R.}~\bibnamefont
  {Jung}}, \bibinfo {author} {\bibfnamefont {I.}~\bibnamefont {Kourakis}},
  \bibinfo {author} {\bibfnamefont {J.}~\bibnamefont {Osterholz}}, \bibinfo
  {author} {\bibfnamefont {F.}~\bibnamefont {Pegoraro}}, \bibinfo {author}
  {\bibfnamefont {A.~P.~L.}\ \bibnamefont {Robinson}}, \bibinfo {author}
  {\bibfnamefont {L.}~\bibnamefont {Romagnani}}, \bibinfo {author}
  {\bibfnamefont {O.}~\bibnamefont {Willi}},\ and\ \bibinfo {author}
  {\bibfnamefont {M.}~\bibnamefont {Borghesi}},\ }\bibfield  {title} {\bibinfo
  {title} {Dynamics of self-generated, large amplitude magnetic fields
  following high-intensity laser matter interaction},\ }\href@noop {}
  {\bibfield  {journal} {\bibinfo  {journal} {Phys. Rev. Lett.}\ }\textbf
  {\bibinfo {volume} {109}},\ \bibinfo {pages} {205002} (\bibinfo {year}
  {2012})}\BibitemShut {NoStop}%
\bibitem [{\citenamefont {Schumaker}\ \emph {et~al.}(2013)\citenamefont
  {Schumaker}, \citenamefont {Nakanii}, \citenamefont {McGuffey}, \citenamefont
  {Zulick}, \citenamefont {Chyvkov}, \citenamefont {Dollar}, \citenamefont
  {Habara}, \citenamefont {Kalintchenko}, \citenamefont {Maksimchuk},
  \citenamefont {Tanaka}, \citenamefont {Thomas}, \citenamefont {Yanovsky},\
  and\ \citenamefont {Krushelnick}}]{B_gen_PRL2013_schumaker}%
  \BibitemOpen
  \bibfield  {author} {\bibinfo {author} {\bibfnamefont {W.}~\bibnamefont
  {Schumaker}}, \bibinfo {author} {\bibfnamefont {N.}~\bibnamefont {Nakanii}},
  \bibinfo {author} {\bibfnamefont {C.}~\bibnamefont {McGuffey}}, \bibinfo
  {author} {\bibfnamefont {C.}~\bibnamefont {Zulick}}, \bibinfo {author}
  {\bibfnamefont {V.}~\bibnamefont {Chyvkov}}, \bibinfo {author} {\bibfnamefont
  {F.}~\bibnamefont {Dollar}}, \bibinfo {author} {\bibfnamefont
  {H.}~\bibnamefont {Habara}}, \bibinfo {author} {\bibfnamefont
  {G.}~\bibnamefont {Kalintchenko}}, \bibinfo {author} {\bibfnamefont
  {A.}~\bibnamefont {Maksimchuk}}, \bibinfo {author} {\bibfnamefont {K.~A.}\
  \bibnamefont {Tanaka}}, \bibinfo {author} {\bibfnamefont {A.~G.~R.}\
  \bibnamefont {Thomas}}, \bibinfo {author} {\bibfnamefont {V.}~\bibnamefont
  {Yanovsky}},\ and\ \bibinfo {author} {\bibfnamefont {K.}~\bibnamefont
  {Krushelnick}},\ }\bibfield  {title} {\bibinfo {title} {Ultrafast electron
  radiography of magnetic fields in high-intensity laser-solid interactions},\
  }\href@noop {} {\bibfield  {journal} {\bibinfo  {journal} {Phys. Rev. Lett.}\
  }\textbf {\bibinfo {volume} {110}},\ \bibinfo {pages} {015003} (\bibinfo
  {year} {2013})}\BibitemShut {NoStop}%
\bibitem [{\citenamefont {Meinecke}\ \emph {et~al.}(2014)\citenamefont
  {Meinecke}, \citenamefont {Doyle}, \citenamefont {Miniati}, \citenamefont
  {Bell}, \citenamefont {Bingham}, \citenamefont {Crowston}, \citenamefont
  {Drake}, \citenamefont {Fatenejad}, \citenamefont {Koenig}, \citenamefont
  {Kuramitsu} \emph {et~al.}}]{turbulent_NP2014_meinecke}%
  \BibitemOpen
  \bibfield  {author} {\bibinfo {author} {\bibfnamefont {J.}~\bibnamefont
  {Meinecke}}, \bibinfo {author} {\bibfnamefont {H.}~\bibnamefont {Doyle}},
  \bibinfo {author} {\bibfnamefont {F.}~\bibnamefont {Miniati}}, \bibinfo
  {author} {\bibfnamefont {A.~R.}\ \bibnamefont {Bell}}, \bibinfo {author}
  {\bibfnamefont {R.}~\bibnamefont {Bingham}}, \bibinfo {author} {\bibfnamefont
  {R.}~\bibnamefont {Crowston}}, \bibinfo {author} {\bibfnamefont
  {R.}~\bibnamefont {Drake}}, \bibinfo {author} {\bibfnamefont
  {M.}~\bibnamefont {Fatenejad}}, \bibinfo {author} {\bibfnamefont
  {M.}~\bibnamefont {Koenig}}, \bibinfo {author} {\bibfnamefont
  {Y.}~\bibnamefont {Kuramitsu}}, \emph {et~al.},\ }\bibfield  {title}
  {\bibinfo {title} {Turbulent amplification of magnetic fields in laboratory
  laser-produced shock waves},\ }\href@noop {} {\bibfield  {journal} {\bibinfo
  {journal} {Nature Physics}\ }\textbf {\bibinfo {volume} {10}},\ \bibinfo
  {pages} {520} (\bibinfo {year} {2014})}\BibitemShut {NoStop}%
\bibitem [{\citenamefont {Huntington}\ \emph {et~al.}(2015)\citenamefont
  {Huntington}, \citenamefont {Fiuza}, \citenamefont {Ross}, \citenamefont
  {Zylstra}, \citenamefont {Drake}, \citenamefont {Froula}, \citenamefont
  {Gregori}, \citenamefont {Kugland}, \citenamefont {Kuranz}, \citenamefont
  {Levy} \emph {et~al.}}]{Weibel_NP2015_huntington}%
  \BibitemOpen
  \bibfield  {author} {\bibinfo {author} {\bibfnamefont {C.}~\bibnamefont
  {Huntington}}, \bibinfo {author} {\bibfnamefont {F.}~\bibnamefont {Fiuza}},
  \bibinfo {author} {\bibfnamefont {J.}~\bibnamefont {Ross}}, \bibinfo {author}
  {\bibfnamefont {A.}~\bibnamefont {Zylstra}}, \bibinfo {author} {\bibfnamefont
  {R.}~\bibnamefont {Drake}}, \bibinfo {author} {\bibfnamefont
  {D.}~\bibnamefont {Froula}}, \bibinfo {author} {\bibfnamefont
  {G.}~\bibnamefont {Gregori}}, \bibinfo {author} {\bibfnamefont
  {N.}~\bibnamefont {Kugland}}, \bibinfo {author} {\bibfnamefont
  {C.}~\bibnamefont {Kuranz}}, \bibinfo {author} {\bibfnamefont
  {M.}~\bibnamefont {Levy}}, \emph {et~al.},\ }\bibfield  {title} {\bibinfo
  {title} {Observation of magnetic field generation via the weibel instability
  in interpenetrating plasma flows},\ }\href@noop {} {\bibfield  {journal}
  {\bibinfo  {journal} {Nature Physics}\ }\textbf {\bibinfo {volume} {11}},\
  \bibinfo {pages} {173} (\bibinfo {year} {2015})}\BibitemShut {NoStop}%
\bibitem [{\citenamefont {Albertazzi}\ \emph {et~al.}(2014)\citenamefont
  {Albertazzi}, \citenamefont {Ciardi}, \citenamefont {Nakatsutsumi},
  \citenamefont {Vinci}, \citenamefont {B{\'e}ard}, \citenamefont {Bonito},
  \citenamefont {Billette}, \citenamefont {Borghesi}, \citenamefont {Burkley},
  \citenamefont {Chen} \emph {et~al.}}]{albertazzi2014laboratory_sci}%
  \BibitemOpen
  \bibfield  {author} {\bibinfo {author} {\bibfnamefont {B.}~\bibnamefont
  {Albertazzi}}, \bibinfo {author} {\bibfnamefont {A.}~\bibnamefont {Ciardi}},
  \bibinfo {author} {\bibfnamefont {M.}~\bibnamefont {Nakatsutsumi}}, \bibinfo
  {author} {\bibfnamefont {T.}~\bibnamefont {Vinci}}, \bibinfo {author}
  {\bibfnamefont {J.}~\bibnamefont {B{\'e}ard}}, \bibinfo {author}
  {\bibfnamefont {R.}~\bibnamefont {Bonito}}, \bibinfo {author} {\bibfnamefont
  {J.}~\bibnamefont {Billette}}, \bibinfo {author} {\bibfnamefont
  {M.}~\bibnamefont {Borghesi}}, \bibinfo {author} {\bibfnamefont
  {Z.}~\bibnamefont {Burkley}}, \bibinfo {author} {\bibfnamefont
  {S.}~\bibnamefont {Chen}}, \emph {et~al.},\ }\bibfield  {title} {\bibinfo
  {title} {Laboratory formation of a scaled protostellar jet by coaligned
  poloidal magnetic field},\ }\href@noop {} {\bibfield  {journal} {\bibinfo
  {journal} {Science}\ }\textbf {\bibinfo {volume} {346}},\ \bibinfo {pages}
  {325} (\bibinfo {year} {2014})}\BibitemShut {NoStop}%
\bibitem [{\citenamefont {Revet}\ \emph {et~al.}(2017)\citenamefont {Revet},
  \citenamefont {Chen}, \citenamefont {Bonito}, \citenamefont {Khiar},
  \citenamefont {Filippov}, \citenamefont {Argiroffi}, \citenamefont
  {Higginson}, \citenamefont {Orlando}, \citenamefont {B{\'e}ard},
  \citenamefont {Blecher} \emph {et~al.}}]{revet2017laboratory_sci_adv}%
  \BibitemOpen
  \bibfield  {author} {\bibinfo {author} {\bibfnamefont {G.}~\bibnamefont
  {Revet}}, \bibinfo {author} {\bibfnamefont {S.~N.}\ \bibnamefont {Chen}},
  \bibinfo {author} {\bibfnamefont {R.}~\bibnamefont {Bonito}}, \bibinfo
  {author} {\bibfnamefont {B.}~\bibnamefont {Khiar}}, \bibinfo {author}
  {\bibfnamefont {E.}~\bibnamefont {Filippov}}, \bibinfo {author}
  {\bibfnamefont {C.}~\bibnamefont {Argiroffi}}, \bibinfo {author}
  {\bibfnamefont {D.~P.}\ \bibnamefont {Higginson}}, \bibinfo {author}
  {\bibfnamefont {S.}~\bibnamefont {Orlando}}, \bibinfo {author} {\bibfnamefont
  {J.}~\bibnamefont {B{\'e}ard}}, \bibinfo {author} {\bibfnamefont
  {M.}~\bibnamefont {Blecher}}, \emph {et~al.},\ }\bibfield  {title} {\bibinfo
  {title} {Laboratory unraveling of matter accretion in young stars},\
  }\href@noop {} {\bibfield  {journal} {\bibinfo  {journal} {Science advances}\
  }\textbf {\bibinfo {volume} {3}},\ \bibinfo {pages} {e1700982} (\bibinfo
  {year} {2017})}\BibitemShut {NoStop}%
\bibitem [{\citenamefont {Revet}\ \emph {et~al.}(2021)\citenamefont {Revet},
  \citenamefont {Khiar}, \citenamefont {Filippov}, \citenamefont {Argiroffi},
  \citenamefont {B{\'e}ard}, \citenamefont {Bonito}, \citenamefont {Cerchez},
  \citenamefont {Chen}, \citenamefont {Gangolf}, \citenamefont {Higginson}
  \emph {et~al.}}]{revet2021laboratory_nc}%
  \BibitemOpen
  \bibfield  {author} {\bibinfo {author} {\bibfnamefont {G.}~\bibnamefont
  {Revet}}, \bibinfo {author} {\bibfnamefont {B.}~\bibnamefont {Khiar}},
  \bibinfo {author} {\bibfnamefont {E.}~\bibnamefont {Filippov}}, \bibinfo
  {author} {\bibfnamefont {C.}~\bibnamefont {Argiroffi}}, \bibinfo {author}
  {\bibfnamefont {J.}~\bibnamefont {B{\'e}ard}}, \bibinfo {author}
  {\bibfnamefont {R.}~\bibnamefont {Bonito}}, \bibinfo {author} {\bibfnamefont
  {M.}~\bibnamefont {Cerchez}}, \bibinfo {author} {\bibfnamefont
  {S.}~\bibnamefont {Chen}}, \bibinfo {author} {\bibfnamefont {T.}~\bibnamefont
  {Gangolf}}, \bibinfo {author} {\bibfnamefont {D.}~\bibnamefont {Higginson}},
  \emph {et~al.},\ }\bibfield  {title} {\bibinfo {title} {Laboratory disruption
  of scaled astrophysical outflows by a misaligned magnetic field},\
  }\href@noop {} {\bibfield  {journal} {\bibinfo  {journal} {Nature
  communications}\ }\textbf {\bibinfo {volume} {12}},\ \bibinfo {pages} {1}
  (\bibinfo {year} {2021})}\BibitemShut {NoStop}%
\bibitem [{\citenamefont {{The Vulcan facility}}()}]{Vulcan10}%
  \BibitemOpen
  \bibfield  {author} {\bibinfo {author} {\bibnamefont {{The Vulcan
  facility}}},\ }\href@noop {} {}\bibinfo {howpublished} {\url{https://www.clf.
  stfc.ac.uk/Pages/The-Vulcan-10-Petawatt-Project.aspx.}}\BibitemShut {Stop}%
\bibitem [{\citenamefont {{The Extreme Light Infrastructure (ELI)}}()}]{ELI}%
  \BibitemOpen
  \bibfield  {author} {\bibinfo {author} {\bibnamefont {{The Extreme Light
  Infrastructure (ELI)}}},\ }\href@noop {} {}\bibinfo {howpublished}
  {\url{http://www.eli-laser.eu/}}\BibitemShut {NoStop}%
\bibitem [{\citenamefont {{ElI-Beamlines}}()}]{ElI-Beamlines}%
  \BibitemOpen
  \bibfield  {author} {\bibinfo {author} {\bibnamefont {{ElI-Beamlines}}},\
  }\href@noop {} {}\bibinfo {howpublished} {\url{https://www.eli-beams.eu/en/
  facility/lasers/}}\BibitemShut {NoStop}%
\bibitem [{\citenamefont {{Exawatt Center for Extreme Light Stidies
  (XCELS)}}()}]{XCELS}%
  \BibitemOpen
  \bibfield  {author} {\bibinfo {author} {\bibnamefont {{Exawatt Center for
  Extreme Light Stidies (XCELS)}}},\ }\href@noop {} {}\bibinfo {howpublished}
  {\url{http://www.xcels.iapras.ru/}}\BibitemShut {NoStop}%
\bibitem [{\citenamefont {Danson}\ \emph {et~al.}(2019)\citenamefont {Danson},
  \citenamefont {Haefner}, \citenamefont {Bromage}, \citenamefont {Butcher},
  \citenamefont {Chanteloup}, \citenamefont {Chowdhury}, \citenamefont
  {Galvanauskas}, \citenamefont {Gizzi}, \citenamefont {Hein}, \citenamefont
  {Hillier} \emph {et~al.}}]{danson2019petawatt}%
  \BibitemOpen
  \bibfield  {author} {\bibinfo {author} {\bibfnamefont {C.~N.}\ \bibnamefont
  {Danson}}, \bibinfo {author} {\bibfnamefont {C.}~\bibnamefont {Haefner}},
  \bibinfo {author} {\bibfnamefont {J.}~\bibnamefont {Bromage}}, \bibinfo
  {author} {\bibfnamefont {T.}~\bibnamefont {Butcher}}, \bibinfo {author}
  {\bibfnamefont {J.-C.~F.}\ \bibnamefont {Chanteloup}}, \bibinfo {author}
  {\bibfnamefont {E.~A.}\ \bibnamefont {Chowdhury}}, \bibinfo {author}
  {\bibfnamefont {A.}~\bibnamefont {Galvanauskas}}, \bibinfo {author}
  {\bibfnamefont {L.~A.}\ \bibnamefont {Gizzi}}, \bibinfo {author}
  {\bibfnamefont {J.}~\bibnamefont {Hein}}, \bibinfo {author} {\bibfnamefont
  {D.~I.}\ \bibnamefont {Hillier}}, \emph {et~al.},\ }\bibfield  {title}
  {\bibinfo {title} {Petawatt and exawatt class lasers worldwide},\ }\href@noop
  {} {\bibfield  {journal} {\bibinfo  {journal} {High Power Laser Science and
  Engineering}\ }\textbf {\bibinfo {volume} {7}} (\bibinfo {year}
  {2019})}\BibitemShut {NoStop}%
\bibitem [{\citenamefont {Yoon}\ \emph {et~al.}(2019)\citenamefont {Yoon},
  \citenamefont {Jeon}, \citenamefont {Shin}, \citenamefont {Lee},
  \citenamefont {Lee}, \citenamefont {Choi}, \citenamefont {Kim}, \citenamefont
  {Sung},\ and\ \citenamefont {Nam}}]{yoon2019achieving}%
  \BibitemOpen
  \bibfield  {author} {\bibinfo {author} {\bibfnamefont {J.~W.}\ \bibnamefont
  {Yoon}}, \bibinfo {author} {\bibfnamefont {C.}~\bibnamefont {Jeon}}, \bibinfo
  {author} {\bibfnamefont {J.}~\bibnamefont {Shin}}, \bibinfo {author}
  {\bibfnamefont {S.~K.}\ \bibnamefont {Lee}}, \bibinfo {author} {\bibfnamefont
  {H.~W.}\ \bibnamefont {Lee}}, \bibinfo {author} {\bibfnamefont {I.~W.}\
  \bibnamefont {Choi}}, \bibinfo {author} {\bibfnamefont {H.~T.}\ \bibnamefont
  {Kim}}, \bibinfo {author} {\bibfnamefont {J.~H.}\ \bibnamefont {Sung}},\ and\
  \bibinfo {author} {\bibfnamefont {C.~H.}\ \bibnamefont {Nam}},\ }\bibfield
  {title} {\bibinfo {title} {Achieving the laser intensity of 5.5$\times$ 10 22
  w/cm 2 with a wavefront-corrected multi-pw laser},\ }\href@noop {} {\bibfield
   {journal} {\bibinfo  {journal} {Optics express}\ }\textbf {\bibinfo {volume}
  {27}},\ \bibinfo {pages} {20412} (\bibinfo {year} {2019})}\BibitemShut
  {NoStop}%
\bibitem [{\citenamefont {Yoon}\ \emph {et~al.}(2021)\citenamefont {Yoon},
  \citenamefont {Kim}, \citenamefont {Choi}, \citenamefont {Sung},
  \citenamefont {Lee}, \citenamefont {Lee},\ and\ \citenamefont
  {Nam}}]{yoon2021realization}%
  \BibitemOpen
  \bibfield  {author} {\bibinfo {author} {\bibfnamefont {J.~W.}\ \bibnamefont
  {Yoon}}, \bibinfo {author} {\bibfnamefont {Y.~G.}\ \bibnamefont {Kim}},
  \bibinfo {author} {\bibfnamefont {I.~W.}\ \bibnamefont {Choi}}, \bibinfo
  {author} {\bibfnamefont {J.~H.}\ \bibnamefont {Sung}}, \bibinfo {author}
  {\bibfnamefont {H.~W.}\ \bibnamefont {Lee}}, \bibinfo {author} {\bibfnamefont
  {S.~K.}\ \bibnamefont {Lee}},\ and\ \bibinfo {author} {\bibfnamefont {C.~H.}\
  \bibnamefont {Nam}},\ }\bibfield  {title} {\bibinfo {title} {Realization of
  laser intensity over 10 23 w/cm 2},\ }\href@noop {} {\bibfield  {journal}
  {\bibinfo  {journal} {Optica}\ }\textbf {\bibinfo {volume} {8}},\ \bibinfo
  {pages} {630} (\bibinfo {year} {2021})}\BibitemShut {NoStop}%
\bibitem [{\citenamefont {Mourou}\ \emph {et~al.}(2006)\citenamefont {Mourou},
  \citenamefont {Tajima},\ and\ \citenamefont {Bulanov}}]{mourou2006optics}%
  \BibitemOpen
  \bibfield  {author} {\bibinfo {author} {\bibfnamefont {G.~A.}\ \bibnamefont
  {Mourou}}, \bibinfo {author} {\bibfnamefont {T.}~\bibnamefont {Tajima}},\
  and\ \bibinfo {author} {\bibfnamefont {S.~V.}\ \bibnamefont {Bulanov}},\
  }\bibfield  {title} {\bibinfo {title} {Optics in the relativistic regime},\
  }\href@noop {} {\bibfield  {journal} {\bibinfo  {journal} {Reviews of modern
  physics}\ }\textbf {\bibinfo {volume} {78}},\ \bibinfo {pages} {309}
  (\bibinfo {year} {2006})}\BibitemShut {NoStop}%
\bibitem [{\citenamefont {Marklund}\ and\ \citenamefont
  {Shukla}(2006)}]{marklund2006nonlinear}%
  \BibitemOpen
  \bibfield  {author} {\bibinfo {author} {\bibfnamefont {M.}~\bibnamefont
  {Marklund}}\ and\ \bibinfo {author} {\bibfnamefont {P.~K.}\ \bibnamefont
  {Shukla}},\ }\bibfield  {title} {\bibinfo {title} {Nonlinear collective
  effects in photon-photon and photon-plasma interactions},\ }\href@noop {}
  {\bibfield  {journal} {\bibinfo  {journal} {Reviews of modern physics}\
  }\textbf {\bibinfo {volume} {78}},\ \bibinfo {pages} {591} (\bibinfo {year}
  {2006})}\BibitemShut {NoStop}%
\bibitem [{\citenamefont {Bell}\ and\ \citenamefont
  {Kirk}(2008)}]{bell2008possibility}%
  \BibitemOpen
  \bibfield  {author} {\bibinfo {author} {\bibfnamefont {A.}~\bibnamefont
  {Bell}}\ and\ \bibinfo {author} {\bibfnamefont {J.~G.}\ \bibnamefont
  {Kirk}},\ }\bibfield  {title} {\bibinfo {title} {Possibility of prolific pair
  production with high-power lasers},\ }\href@noop {} {\bibfield  {journal}
  {\bibinfo  {journal} {Physical review letters}\ }\textbf {\bibinfo {volume}
  {101}},\ \bibinfo {pages} {200403} (\bibinfo {year} {2008})}\BibitemShut
  {NoStop}%
\bibitem [{\citenamefont {Di~Piazza}\ \emph {et~al.}(2012)\citenamefont
  {Di~Piazza}, \citenamefont {M{\"u}ller}, \citenamefont {Hatsagortsyan},\ and\
  \citenamefont {Keitel}}]{di2012extremely}%
  \BibitemOpen
  \bibfield  {author} {\bibinfo {author} {\bibfnamefont {A.}~\bibnamefont
  {Di~Piazza}}, \bibinfo {author} {\bibfnamefont {C.}~\bibnamefont
  {M{\"u}ller}}, \bibinfo {author} {\bibfnamefont {K.}~\bibnamefont
  {Hatsagortsyan}},\ and\ \bibinfo {author} {\bibfnamefont {C.~H.}\
  \bibnamefont {Keitel}},\ }\bibfield  {title} {\bibinfo {title} {Extremely
  high-intensity laser interactions with fundamental quantum systems},\
  }\href@noop {} {\bibfield  {journal} {\bibinfo  {journal} {Reviews of Modern
  Physics}\ }\textbf {\bibinfo {volume} {84}},\ \bibinfo {pages} {1177}
  (\bibinfo {year} {2012})}\BibitemShut {NoStop}%
\bibitem [{\citenamefont {Bulanov}\ \emph {et~al.}(2015)\citenamefont
  {Bulanov}, \citenamefont {Esirkepov}, \citenamefont {Kando}, \citenamefont
  {Koga}, \citenamefont {Kondo},\ and\ \citenamefont {Korn}}]{RRD_bulanov2015}%
  \BibitemOpen
  \bibfield  {author} {\bibinfo {author} {\bibfnamefont {S.}~\bibnamefont
  {Bulanov}}, \bibinfo {author} {\bibfnamefont {T.~Z.}\ \bibnamefont
  {Esirkepov}}, \bibinfo {author} {\bibfnamefont {M.}~\bibnamefont {Kando}},
  \bibinfo {author} {\bibfnamefont {J.}~\bibnamefont {Koga}}, \bibinfo {author}
  {\bibfnamefont {K.}~\bibnamefont {Kondo}},\ and\ \bibinfo {author}
  {\bibfnamefont {G.}~\bibnamefont {Korn}},\ }\bibfield  {title} {\bibinfo
  {title} {On the problems of relativistic laboratory astrophysics and
  fundamental physics with super powerful lasers},\ }\href@noop {} {\bibfield
  {journal} {\bibinfo  {journal} {Plasma Physics Reports}\ }\textbf {\bibinfo
  {volume} {41}},\ \bibinfo {pages} {1} (\bibinfo {year} {2015})}\BibitemShut
  {NoStop}%
\bibitem [{\citenamefont {Poder}\ \emph {et~al.}(2018)\citenamefont {Poder},
  \citenamefont {Tamburini}, \citenamefont {Sarri}, \citenamefont {Di~Piazza},
  \citenamefont {Kuschel}, \citenamefont {Baird}, \citenamefont {Behm},
  \citenamefont {Bohlen}, \citenamefont {Cole}, \citenamefont {Corvan} \emph
  {et~al.}}]{poder2018experimental}%
  \BibitemOpen
  \bibfield  {author} {\bibinfo {author} {\bibfnamefont {K.}~\bibnamefont
  {Poder}}, \bibinfo {author} {\bibfnamefont {M.}~\bibnamefont {Tamburini}},
  \bibinfo {author} {\bibfnamefont {G.}~\bibnamefont {Sarri}}, \bibinfo
  {author} {\bibfnamefont {A.}~\bibnamefont {Di~Piazza}}, \bibinfo {author}
  {\bibfnamefont {S.}~\bibnamefont {Kuschel}}, \bibinfo {author} {\bibfnamefont
  {C.}~\bibnamefont {Baird}}, \bibinfo {author} {\bibfnamefont
  {K.}~\bibnamefont {Behm}}, \bibinfo {author} {\bibfnamefont {S.}~\bibnamefont
  {Bohlen}}, \bibinfo {author} {\bibfnamefont {J.}~\bibnamefont {Cole}},
  \bibinfo {author} {\bibfnamefont {D.}~\bibnamefont {Corvan}}, \emph
  {et~al.},\ }\bibfield  {title} {\bibinfo {title} {Experimental signatures of
  the quantum nature of radiation reaction in the field of an ultraintense
  laser},\ }\href@noop {} {\bibfield  {journal} {\bibinfo  {journal} {Physical
  Review X}\ }\textbf {\bibinfo {volume} {8}},\ \bibinfo {pages} {031004}
  (\bibinfo {year} {2018})}\BibitemShut {NoStop}%
\bibitem [{\citenamefont {Cole}\ \emph {et~al.}(2018)\citenamefont {Cole},
  \citenamefont {Behm}, \citenamefont {Gerstmayr}, \citenamefont {Blackburn},
  \citenamefont {Wood}, \citenamefont {Baird}, \citenamefont {Duff},
  \citenamefont {Harvey}, \citenamefont {Ilderton}, \citenamefont {Joglekar}
  \emph {et~al.}}]{cole2018experimental}%
  \BibitemOpen
  \bibfield  {author} {\bibinfo {author} {\bibfnamefont {J.}~\bibnamefont
  {Cole}}, \bibinfo {author} {\bibfnamefont {K.}~\bibnamefont {Behm}}, \bibinfo
  {author} {\bibfnamefont {E.}~\bibnamefont {Gerstmayr}}, \bibinfo {author}
  {\bibfnamefont {T.}~\bibnamefont {Blackburn}}, \bibinfo {author}
  {\bibfnamefont {J.}~\bibnamefont {Wood}}, \bibinfo {author} {\bibfnamefont
  {C.}~\bibnamefont {Baird}}, \bibinfo {author} {\bibfnamefont {M.~J.}\
  \bibnamefont {Duff}}, \bibinfo {author} {\bibfnamefont {C.}~\bibnamefont
  {Harvey}}, \bibinfo {author} {\bibfnamefont {A.}~\bibnamefont {Ilderton}},
  \bibinfo {author} {\bibfnamefont {A.}~\bibnamefont {Joglekar}}, \emph
  {et~al.},\ }\bibfield  {title} {\bibinfo {title} {Experimental evidence of
  radiation reaction in the collision of a high-intensity laser pulse with a
  laser-wakefield accelerated electron beam},\ }\href@noop {} {\bibfield
  {journal} {\bibinfo  {journal} {Physical Review X}\ }\textbf {\bibinfo
  {volume} {8}},\ \bibinfo {pages} {011020} (\bibinfo {year}
  {2018})}\BibitemShut {NoStop}%
\bibitem [{\citenamefont {Li}\ \emph {et~al.}(2007)\citenamefont {Li},
  \citenamefont {S{\'e}guin}, \citenamefont {Frenje}, \citenamefont {Rygg},
  \citenamefont {Petrasso}, \citenamefont {Town}, \citenamefont {Amendt},
  \citenamefont {Hatchett}, \citenamefont {Landen}, \citenamefont {Mackinnon}
  \emph {et~al.}}]{detect_li2007observation}%
  \BibitemOpen
  \bibfield  {author} {\bibinfo {author} {\bibfnamefont {C.}~\bibnamefont
  {Li}}, \bibinfo {author} {\bibfnamefont {F.}~\bibnamefont {S{\'e}guin}},
  \bibinfo {author} {\bibfnamefont {J.}~\bibnamefont {Frenje}}, \bibinfo
  {author} {\bibfnamefont {J.}~\bibnamefont {Rygg}}, \bibinfo {author}
  {\bibfnamefont {R.}~\bibnamefont {Petrasso}}, \bibinfo {author}
  {\bibfnamefont {R.}~\bibnamefont {Town}}, \bibinfo {author} {\bibfnamefont
  {P.}~\bibnamefont {Amendt}}, \bibinfo {author} {\bibfnamefont
  {S.}~\bibnamefont {Hatchett}}, \bibinfo {author} {\bibfnamefont
  {O.}~\bibnamefont {Landen}}, \bibinfo {author} {\bibfnamefont
  {A.}~\bibnamefont {Mackinnon}}, \emph {et~al.},\ }\bibfield  {title}
  {\bibinfo {title} {Observation of the decay dynamics and instabilities of
  megagauss field structures in laser-produced plasmas},\ }\href@noop {}
  {\bibfield  {journal} {\bibinfo  {journal} {Physical review letters}\
  }\textbf {\bibinfo {volume} {99}},\ \bibinfo {pages} {015001} (\bibinfo
  {year} {2007})}\BibitemShut {NoStop}%
\bibitem [{\citenamefont {Willingale}\ \emph {et~al.}(2010)\citenamefont
  {Willingale}, \citenamefont {Thomas}, \citenamefont {Nilson}, \citenamefont
  {Kaluza}, \citenamefont {Bandyopadhyay}, \citenamefont {Dangor},
  \citenamefont {Evans}, \citenamefont {Fernandes}, \citenamefont {Haines},
  \citenamefont {Kamperidis} \emph {et~al.}}]{detect_willingale2010fast}%
  \BibitemOpen
  \bibfield  {author} {\bibinfo {author} {\bibfnamefont {L.}~\bibnamefont
  {Willingale}}, \bibinfo {author} {\bibfnamefont {A.}~\bibnamefont {Thomas}},
  \bibinfo {author} {\bibfnamefont {P.}~\bibnamefont {Nilson}}, \bibinfo
  {author} {\bibfnamefont {M.}~\bibnamefont {Kaluza}}, \bibinfo {author}
  {\bibfnamefont {S.}~\bibnamefont {Bandyopadhyay}}, \bibinfo {author}
  {\bibfnamefont {A.}~\bibnamefont {Dangor}}, \bibinfo {author} {\bibfnamefont
  {R.}~\bibnamefont {Evans}}, \bibinfo {author} {\bibfnamefont
  {P.}~\bibnamefont {Fernandes}}, \bibinfo {author} {\bibfnamefont
  {M.}~\bibnamefont {Haines}}, \bibinfo {author} {\bibfnamefont
  {C.}~\bibnamefont {Kamperidis}}, \emph {et~al.},\ }\bibfield  {title}
  {\bibinfo {title} {Fast advection of magnetic fields by hot electrons},\
  }\href@noop {} {\bibfield  {journal} {\bibinfo  {journal} {Physical review
  letters}\ }\textbf {\bibinfo {volume} {105}},\ \bibinfo {pages} {095001}
  (\bibinfo {year} {2010})}\BibitemShut {NoStop}%
\bibitem [{\citenamefont {Macchi}\ \emph {et~al.}(2013)\citenamefont {Macchi},
  \citenamefont {Borghesi},\ and\ \citenamefont
  {Passoni}}]{detect_macchi2013ion}%
  \BibitemOpen
  \bibfield  {author} {\bibinfo {author} {\bibfnamefont {A.}~\bibnamefont
  {Macchi}}, \bibinfo {author} {\bibfnamefont {M.}~\bibnamefont {Borghesi}},\
  and\ \bibinfo {author} {\bibfnamefont {M.}~\bibnamefont {Passoni}},\
  }\bibfield  {title} {\bibinfo {title} {Ion acceleration by superintense
  laser-plasma interaction},\ }\href@noop {} {\bibfield  {journal} {\bibinfo
  {journal} {Reviews of Modern Physics}\ }\textbf {\bibinfo {volume} {85}},\
  \bibinfo {pages} {751} (\bibinfo {year} {2013})}\BibitemShut {NoStop}%
\bibitem [{\citenamefont {Lancia}\ \emph {et~al.}(2014)\citenamefont {Lancia},
  \citenamefont {Albertazzi}, \citenamefont {Boniface}, \citenamefont
  {Grisollet}, \citenamefont {Riquier}, \citenamefont {Chaland}, \citenamefont
  {Le~Thanh}, \citenamefont {Mellor}, \citenamefont {Antici}, \citenamefont
  {Buffechoux} \emph {et~al.}}]{detect_lancia2014topology}%
  \BibitemOpen
  \bibfield  {author} {\bibinfo {author} {\bibfnamefont {L.}~\bibnamefont
  {Lancia}}, \bibinfo {author} {\bibfnamefont {B.}~\bibnamefont {Albertazzi}},
  \bibinfo {author} {\bibfnamefont {C.}~\bibnamefont {Boniface}}, \bibinfo
  {author} {\bibfnamefont {A.}~\bibnamefont {Grisollet}}, \bibinfo {author}
  {\bibfnamefont {R.}~\bibnamefont {Riquier}}, \bibinfo {author} {\bibfnamefont
  {F.}~\bibnamefont {Chaland}}, \bibinfo {author} {\bibfnamefont {K.-C.}\
  \bibnamefont {Le~Thanh}}, \bibinfo {author} {\bibfnamefont {P.}~\bibnamefont
  {Mellor}}, \bibinfo {author} {\bibfnamefont {P.}~\bibnamefont {Antici}},
  \bibinfo {author} {\bibfnamefont {S.}~\bibnamefont {Buffechoux}}, \emph
  {et~al.},\ }\bibfield  {title} {\bibinfo {title} {Topology of megagauss
  magnetic fields and of heat-carrying electrons produced in a high-power
  laser-solid interaction},\ }\href@noop {} {\bibfield  {journal} {\bibinfo
  {journal} {Physical review letters}\ }\textbf {\bibinfo {volume} {113}},\
  \bibinfo {pages} {235001} (\bibinfo {year} {2014})}\BibitemShut {NoStop}%
\bibitem [{\citenamefont {Zhang}\ \emph
  {et~al.}(2020{\natexlab{a}})\citenamefont {Zhang}, \citenamefont {Hua},
  \citenamefont {Wu}, \citenamefont {Fang}, \citenamefont {Ma}, \citenamefont
  {Zhang}, \citenamefont {Liu}, \citenamefont {Peng}, \citenamefont {He},
  \citenamefont {Huang} \emph {et~al.}}]{detect_zhang2020measurements}%
  \BibitemOpen
  \bibfield  {author} {\bibinfo {author} {\bibfnamefont {C.}~\bibnamefont
  {Zhang}}, \bibinfo {author} {\bibfnamefont {J.}~\bibnamefont {Hua}}, \bibinfo
  {author} {\bibfnamefont {Y.}~\bibnamefont {Wu}}, \bibinfo {author}
  {\bibfnamefont {Y.}~\bibnamefont {Fang}}, \bibinfo {author} {\bibfnamefont
  {Y.}~\bibnamefont {Ma}}, \bibinfo {author} {\bibfnamefont {T.}~\bibnamefont
  {Zhang}}, \bibinfo {author} {\bibfnamefont {S.}~\bibnamefont {Liu}}, \bibinfo
  {author} {\bibfnamefont {B.}~\bibnamefont {Peng}}, \bibinfo {author}
  {\bibfnamefont {Y.}~\bibnamefont {He}}, \bibinfo {author} {\bibfnamefont
  {C.-K.}\ \bibnamefont {Huang}}, \emph {et~al.},\ }\bibfield  {title}
  {\bibinfo {title} {Measurements of the growth and saturation of electron
  weibel instability in optical-field ionized plasmas},\ }\href@noop {}
  {\bibfield  {journal} {\bibinfo  {journal} {Physical Review Letters}\
  }\textbf {\bibinfo {volume} {125}},\ \bibinfo {pages} {255001} (\bibinfo
  {year} {2020}{\natexlab{a}})}\BibitemShut {NoStop}%
\bibitem [{\citenamefont {Bott}\ \emph {et~al.}(2021)\citenamefont {Bott},
  \citenamefont {Tzeferacos}, \citenamefont {Chen}, \citenamefont {Palmer},
  \citenamefont {Rigby}, \citenamefont {Bell}, \citenamefont {Bingham},
  \citenamefont {Birkel}, \citenamefont {Graziani}, \citenamefont {Froula}
  \emph {et~al.}}]{proton_imaging_bott2021time}%
  \BibitemOpen
  \bibfield  {author} {\bibinfo {author} {\bibfnamefont {A.~F.}\ \bibnamefont
  {Bott}}, \bibinfo {author} {\bibfnamefont {P.}~\bibnamefont {Tzeferacos}},
  \bibinfo {author} {\bibfnamefont {L.}~\bibnamefont {Chen}}, \bibinfo {author}
  {\bibfnamefont {C.~A.}\ \bibnamefont {Palmer}}, \bibinfo {author}
  {\bibfnamefont {A.}~\bibnamefont {Rigby}}, \bibinfo {author} {\bibfnamefont
  {A.~R.}\ \bibnamefont {Bell}}, \bibinfo {author} {\bibfnamefont
  {R.}~\bibnamefont {Bingham}}, \bibinfo {author} {\bibfnamefont
  {A.}~\bibnamefont {Birkel}}, \bibinfo {author} {\bibfnamefont
  {C.}~\bibnamefont {Graziani}}, \bibinfo {author} {\bibfnamefont {D.~H.}\
  \bibnamefont {Froula}}, \emph {et~al.},\ }\bibfield  {title} {\bibinfo
  {title} {Time-resolved turbulent dynamo in a laser plasma},\ }\href@noop {}
  {\bibfield  {journal} {\bibinfo  {journal} {Proceedings of the National
  Academy of Sciences}\ }\textbf {\bibinfo {volume} {118}} (\bibinfo {year}
  {2021})}\BibitemShut {NoStop}%
\bibitem [{\citenamefont {Tatarakis}\ \emph
  {et~al.}(2002{\natexlab{b}})\citenamefont {Tatarakis}, \citenamefont {Gopal},
  \citenamefont {Watts}, \citenamefont {Beg}, \citenamefont {Dangor},
  \citenamefont {Krushelnick}, \citenamefont {Wagner}, \citenamefont {Norreys},
  \citenamefont {Clark}, \citenamefont {Zepf} \emph
  {et~al.}}]{FR_tatarakis2002measurements}%
  \BibitemOpen
  \bibfield  {author} {\bibinfo {author} {\bibfnamefont {M.}~\bibnamefont
  {Tatarakis}}, \bibinfo {author} {\bibfnamefont {A.}~\bibnamefont {Gopal}},
  \bibinfo {author} {\bibfnamefont {I.}~\bibnamefont {Watts}}, \bibinfo
  {author} {\bibfnamefont {F.}~\bibnamefont {Beg}}, \bibinfo {author}
  {\bibfnamefont {A.}~\bibnamefont {Dangor}}, \bibinfo {author} {\bibfnamefont
  {K.}~\bibnamefont {Krushelnick}}, \bibinfo {author} {\bibfnamefont
  {U.}~\bibnamefont {Wagner}}, \bibinfo {author} {\bibfnamefont
  {P.}~\bibnamefont {Norreys}}, \bibinfo {author} {\bibfnamefont
  {E.}~\bibnamefont {Clark}}, \bibinfo {author} {\bibfnamefont
  {M.}~\bibnamefont {Zepf}}, \emph {et~al.},\ }\bibfield  {title} {\bibinfo
  {title} {Measurements of ultrastrong magnetic fields during relativistic
  laser--plasma interactions},\ }\href@noop {} {\bibfield  {journal} {\bibinfo
  {journal} {Physics of Plasmas}\ }\textbf {\bibinfo {volume} {9}},\ \bibinfo
  {pages} {2244} (\bibinfo {year} {2002}{\natexlab{b}})}\BibitemShut {NoStop}%
\bibitem [{\citenamefont {Kaluza}\ \emph {et~al.}(2010)\citenamefont {Kaluza},
  \citenamefont {Schlenvoigt}, \citenamefont {Mangles}, \citenamefont {Thomas},
  \citenamefont {Dangor}, \citenamefont {Schwoerer}, \citenamefont {Mori},
  \citenamefont {Najmudin},\ and\ \citenamefont
  {Krushelnick}}]{FR_kaluza2010measurement}%
  \BibitemOpen
  \bibfield  {author} {\bibinfo {author} {\bibfnamefont {M.}~\bibnamefont
  {Kaluza}}, \bibinfo {author} {\bibfnamefont {H.-P.}\ \bibnamefont
  {Schlenvoigt}}, \bibinfo {author} {\bibfnamefont {S.}~\bibnamefont
  {Mangles}}, \bibinfo {author} {\bibfnamefont {A.}~\bibnamefont {Thomas}},
  \bibinfo {author} {\bibfnamefont {A.}~\bibnamefont {Dangor}}, \bibinfo
  {author} {\bibfnamefont {H.}~\bibnamefont {Schwoerer}}, \bibinfo {author}
  {\bibfnamefont {W.}~\bibnamefont {Mori}}, \bibinfo {author} {\bibfnamefont
  {Z.}~\bibnamefont {Najmudin}},\ and\ \bibinfo {author} {\bibfnamefont
  {K.}~\bibnamefont {Krushelnick}},\ }\bibfield  {title} {\bibinfo {title}
  {Measurement of magnetic-field structures in a laser-wakefield accelerator},\
  }\href@noop {} {\bibfield  {journal} {\bibinfo  {journal} {Physical review
  letters}\ }\textbf {\bibinfo {volume} {105}},\ \bibinfo {pages} {115002}
  (\bibinfo {year} {2010})}\BibitemShut {NoStop}%
\bibitem [{\citenamefont {Buck}\ \emph {et~al.}(2011)\citenamefont {Buck},
  \citenamefont {Nicolai}, \citenamefont {Schmid}, \citenamefont {Sears},
  \citenamefont {S{\"a}vert}, \citenamefont {Mikhailova}, \citenamefont
  {Krausz}, \citenamefont {Kaluza},\ and\ \citenamefont
  {Veisz}}]{FR_buck2011real}%
  \BibitemOpen
  \bibfield  {author} {\bibinfo {author} {\bibfnamefont {A.}~\bibnamefont
  {Buck}}, \bibinfo {author} {\bibfnamefont {M.}~\bibnamefont {Nicolai}},
  \bibinfo {author} {\bibfnamefont {K.}~\bibnamefont {Schmid}}, \bibinfo
  {author} {\bibfnamefont {C.~M.}\ \bibnamefont {Sears}}, \bibinfo {author}
  {\bibfnamefont {A.}~\bibnamefont {S{\"a}vert}}, \bibinfo {author}
  {\bibfnamefont {J.~M.}\ \bibnamefont {Mikhailova}}, \bibinfo {author}
  {\bibfnamefont {F.}~\bibnamefont {Krausz}}, \bibinfo {author} {\bibfnamefont
  {M.~C.}\ \bibnamefont {Kaluza}},\ and\ \bibinfo {author} {\bibfnamefont
  {L.}~\bibnamefont {Veisz}},\ }\bibfield  {title} {\bibinfo {title} {Real-time
  observation of laser-driven electron acceleration},\ }\href@noop {}
  {\bibfield  {journal} {\bibinfo  {journal} {Nature Physics}\ }\textbf
  {\bibinfo {volume} {7}},\ \bibinfo {pages} {543} (\bibinfo {year}
  {2011})}\BibitemShut {NoStop}%
\bibitem [{\citenamefont {Zhou}\ \emph {et~al.}(2018)\citenamefont {Zhou},
  \citenamefont {Bai}, \citenamefont {Tian}, \citenamefont {Sun}, \citenamefont
  {Cao},\ and\ \citenamefont {Liu}}]{FR_zhou2018self}%
  \BibitemOpen
  \bibfield  {author} {\bibinfo {author} {\bibfnamefont {S.}~\bibnamefont
  {Zhou}}, \bibinfo {author} {\bibfnamefont {Y.}~\bibnamefont {Bai}}, \bibinfo
  {author} {\bibfnamefont {Y.}~\bibnamefont {Tian}}, \bibinfo {author}
  {\bibfnamefont {H.}~\bibnamefont {Sun}}, \bibinfo {author} {\bibfnamefont
  {L.}~\bibnamefont {Cao}},\ and\ \bibinfo {author} {\bibfnamefont
  {J.}~\bibnamefont {Liu}},\ }\bibfield  {title} {\bibinfo {title}
  {Self-organized kilotesla magnetic-tube array in an expanding spherical
  plasma irradiated by khz femtosecond laser pulses},\ }\href@noop {}
  {\bibfield  {journal} {\bibinfo  {journal} {Physical review letters}\
  }\textbf {\bibinfo {volume} {121}},\ \bibinfo {pages} {255002} (\bibinfo
  {year} {2018})}\BibitemShut {NoStop}%
\bibitem [{\citenamefont {Wang}\ \emph {et~al.}(2019)\citenamefont {Wang},
  \citenamefont {Toncian}, \citenamefont {Wei},\ and\ \citenamefont
  {Arefiev}}]{wang2019structured}%
  \BibitemOpen
  \bibfield  {author} {\bibinfo {author} {\bibfnamefont {T.}~\bibnamefont
  {Wang}}, \bibinfo {author} {\bibfnamefont {T.}~\bibnamefont {Toncian}},
  \bibinfo {author} {\bibfnamefont {M.}~\bibnamefont {Wei}},\ and\ \bibinfo
  {author} {\bibfnamefont {A.}~\bibnamefont {Arefiev}},\ }\bibfield  {title}
  {\bibinfo {title} {Structured targets for detection of megatesla-level
  magnetic fields through faraday rotation of xfel beams},\ }\href@noop {}
  {\bibfield  {journal} {\bibinfo  {journal} {Physics of plasmas}\ }\textbf
  {\bibinfo {volume} {26}},\ \bibinfo {pages} {013105} (\bibinfo {year}
  {2019})}\BibitemShut {NoStop}%
\bibitem [{\citenamefont {Jungwirth}\ \emph {et~al.}(2014)\citenamefont
  {Jungwirth}, \citenamefont {Wunderlich}, \citenamefont {Nov{\'a}k},
  \citenamefont {Olejn{\'\i}k}, \citenamefont {Gallagher}, \citenamefont
  {Campion}, \citenamefont {Edmonds}, \citenamefont {Rushforth}, \citenamefont
  {Ferguson},\ and\ \citenamefont {N{\v{e}}mec}}]{jungwirth2014spin}%
  \BibitemOpen
  \bibfield  {author} {\bibinfo {author} {\bibfnamefont {T.}~\bibnamefont
  {Jungwirth}}, \bibinfo {author} {\bibfnamefont {J.}~\bibnamefont
  {Wunderlich}}, \bibinfo {author} {\bibfnamefont {V.}~\bibnamefont
  {Nov{\'a}k}}, \bibinfo {author} {\bibfnamefont {K.}~\bibnamefont
  {Olejn{\'\i}k}}, \bibinfo {author} {\bibfnamefont {B.}~\bibnamefont
  {Gallagher}}, \bibinfo {author} {\bibfnamefont {R.}~\bibnamefont {Campion}},
  \bibinfo {author} {\bibfnamefont {K.}~\bibnamefont {Edmonds}}, \bibinfo
  {author} {\bibfnamefont {A.}~\bibnamefont {Rushforth}}, \bibinfo {author}
  {\bibfnamefont {A.}~\bibnamefont {Ferguson}},\ and\ \bibinfo {author}
  {\bibfnamefont {P.}~\bibnamefont {N{\v{e}}mec}},\ }\bibfield  {title}
  {\bibinfo {title} {Spin-dependent phenomena and device concepts explored in
  (ga, mn) as},\ }\href@noop {} {\bibfield  {journal} {\bibinfo  {journal}
  {Reviews of Modern Physics}\ }\textbf {\bibinfo {volume} {86}},\ \bibinfo
  {pages} {855} (\bibinfo {year} {2014})}\BibitemShut {NoStop}%
\bibitem [{\citenamefont {Abe}\ \emph {et~al.}(1995)\citenamefont {Abe},
  \citenamefont {Akagi}, \citenamefont {Anthony}, \citenamefont {Antonov},
  \citenamefont {Arnold}, \citenamefont {Averett}, \citenamefont {Band},
  \citenamefont {Bauer}, \citenamefont {Borel}, \citenamefont {Bosted} \emph
  {et~al.}}]{pol_e_nuclear_structure_1}%
  \BibitemOpen
  \bibfield  {author} {\bibinfo {author} {\bibfnamefont {K.}~\bibnamefont
  {Abe}}, \bibinfo {author} {\bibfnamefont {T.}~\bibnamefont {Akagi}}, \bibinfo
  {author} {\bibfnamefont {P.}~\bibnamefont {Anthony}}, \bibinfo {author}
  {\bibfnamefont {R.}~\bibnamefont {Antonov}}, \bibinfo {author} {\bibfnamefont
  {R.}~\bibnamefont {Arnold}}, \bibinfo {author} {\bibfnamefont
  {T.}~\bibnamefont {Averett}}, \bibinfo {author} {\bibfnamefont
  {H.}~\bibnamefont {Band}}, \bibinfo {author} {\bibfnamefont {J.}~\bibnamefont
  {Bauer}}, \bibinfo {author} {\bibfnamefont {H.}~\bibnamefont {Borel}},
  \bibinfo {author} {\bibfnamefont {P.}~\bibnamefont {Bosted}}, \emph
  {et~al.},\ }\bibfield  {title} {\bibinfo {title} {Precision measurement of
  the deuteron spin structure function g 1 d},\ }\href@noop {} {\bibfield
  {journal} {\bibinfo  {journal} {Phys. Rev. Lett.}\ }\textbf {\bibinfo
  {volume} {75}},\ \bibinfo {pages} {25} (\bibinfo {year} {1995})}\BibitemShut
  {NoStop}%
\bibitem [{\citenamefont {Moortgat-Pick}\ \emph {et~al.}(2008)\citenamefont
  {Moortgat-Pick}, \citenamefont {Abe}, \citenamefont {Alexander},
  \citenamefont {Ananthanarayan}, \citenamefont {Babich}, \citenamefont
  {Bharadwaj}, \citenamefont {Barber}, \citenamefont {Bartl}, \citenamefont
  {Brachmann}, \citenamefont {Chen} \emph
  {et~al.}}]{pol_e_beyound_stand_model}%
  \BibitemOpen
  \bibfield  {author} {\bibinfo {author} {\bibfnamefont {G.}~\bibnamefont
  {Moortgat-Pick}}, \bibinfo {author} {\bibfnamefont {T.}~\bibnamefont {Abe}},
  \bibinfo {author} {\bibfnamefont {G.}~\bibnamefont {Alexander}}, \bibinfo
  {author} {\bibfnamefont {B.}~\bibnamefont {Ananthanarayan}}, \bibinfo
  {author} {\bibfnamefont {A.}~\bibnamefont {Babich}}, \bibinfo {author}
  {\bibfnamefont {V.}~\bibnamefont {Bharadwaj}}, \bibinfo {author}
  {\bibfnamefont {D.}~\bibnamefont {Barber}}, \bibinfo {author} {\bibfnamefont
  {A.}~\bibnamefont {Bartl}}, \bibinfo {author} {\bibfnamefont
  {A.}~\bibnamefont {Brachmann}}, \bibinfo {author} {\bibfnamefont
  {S.}~\bibnamefont {Chen}}, \emph {et~al.},\ }\bibfield  {title} {\bibinfo
  {title} {Polarized positrons and electrons at the linear collider},\
  }\href@noop {} {\bibfield  {journal} {\bibinfo  {journal} {Physics Reports}\
  }\textbf {\bibinfo {volume} {460}},\ \bibinfo {pages} {131} (\bibinfo {year}
  {2008})}\BibitemShut {NoStop}%
\bibitem [{\citenamefont {Walser}\ \emph {et~al.}(2002)\citenamefont {Walser},
  \citenamefont {Urbach}, \citenamefont {Hatsagortsyan}, \citenamefont {Hu},\
  and\ \citenamefont {Keitel}}]{Walser_2002}%
  \BibitemOpen
  \bibfield  {author} {\bibinfo {author} {\bibfnamefont {M.~W.}\ \bibnamefont
  {Walser}}, \bibinfo {author} {\bibfnamefont {D.~J.}\ \bibnamefont {Urbach}},
  \bibinfo {author} {\bibfnamefont {K.~Z.}\ \bibnamefont {Hatsagortsyan}},
  \bibinfo {author} {\bibfnamefont {S.~X.}\ \bibnamefont {Hu}},\ and\ \bibinfo
  {author} {\bibfnamefont {C.~H.}\ \bibnamefont {Keitel}},\ }\bibfield  {title}
  {\bibinfo {title} {Spin and radiation in intense laser fields},\ }\href@noop
  {} {\bibfield  {journal} {\bibinfo  {journal} {Phys. Rev. A}\ }\textbf
  {\bibinfo {volume} {65}},\ \bibinfo {pages} {043410} (\bibinfo {year}
  {2002})}\BibitemShut {NoStop}%
\bibitem [{\citenamefont {Del~Sorbo}\ \emph {et~al.}(2017)\citenamefont
  {Del~Sorbo}, \citenamefont {Seipt}, \citenamefont {Blackburn}, \citenamefont
  {Thomas}, \citenamefont {Murphy}, \citenamefont {Kirk},\ and\ \citenamefont
  {Ridgers}}]{Sorbo_2017}%
  \BibitemOpen
  \bibfield  {author} {\bibinfo {author} {\bibfnamefont {D.}~\bibnamefont
  {Del~Sorbo}}, \bibinfo {author} {\bibfnamefont {D.}~\bibnamefont {Seipt}},
  \bibinfo {author} {\bibfnamefont {T.~G.}\ \bibnamefont {Blackburn}}, \bibinfo
  {author} {\bibfnamefont {A.~G.~R.}\ \bibnamefont {Thomas}}, \bibinfo {author}
  {\bibfnamefont {C.~D.}\ \bibnamefont {Murphy}}, \bibinfo {author}
  {\bibfnamefont {J.~G.}\ \bibnamefont {Kirk}},\ and\ \bibinfo {author}
  {\bibfnamefont {C.~P.}\ \bibnamefont {Ridgers}},\ }\bibfield  {title}
  {\bibinfo {title} {Spin polarization of electrons by ultraintense lasers},\
  }\href {https://doi.org/10.1103/PhysRevA.96.043407} {\bibfield  {journal}
  {\bibinfo  {journal} {Phys. Rev. A}\ }\textbf {\bibinfo {volume} {96}},\
  \bibinfo {pages} {043407} (\bibinfo {year} {2017})}\BibitemShut {NoStop}%
\bibitem [{\citenamefont {Seipt}\ \emph {et~al.}(2018)\citenamefont {Seipt},
  \citenamefont {Del~Sorbo}, \citenamefont {Ridgers},\ and\ \citenamefont
  {Thomas}}]{Seipt_2018}%
  \BibitemOpen
  \bibfield  {author} {\bibinfo {author} {\bibfnamefont {D.}~\bibnamefont
  {Seipt}}, \bibinfo {author} {\bibfnamefont {D.}~\bibnamefont {Del~Sorbo}},
  \bibinfo {author} {\bibfnamefont {C.~P.}\ \bibnamefont {Ridgers}},\ and\
  \bibinfo {author} {\bibfnamefont {A.~G.~R.}\ \bibnamefont {Thomas}},\
  }\bibfield  {title} {\bibinfo {title} {Theory of radiative electron
  polarization in strong laser fields},\ }\href@noop {} {\bibfield  {journal}
  {\bibinfo  {journal} {Phys. Rev. A}\ }\textbf {\bibinfo {volume} {98}},\
  \bibinfo {pages} {023417} (\bibinfo {year} {2018})}\BibitemShut {NoStop}%
\bibitem [{\citenamefont {Sorbo}\ \emph {et~al.}(2018)\citenamefont {Sorbo},
  \citenamefont {Seipt}, \citenamefont {Thomas},\ and\ \citenamefont
  {Ridgers}}]{sorbo2018ppcf}%
  \BibitemOpen
  \bibfield  {author} {\bibinfo {author} {\bibfnamefont {D.~D.}\ \bibnamefont
  {Sorbo}}, \bibinfo {author} {\bibfnamefont {D.}~\bibnamefont {Seipt}},
  \bibinfo {author} {\bibfnamefont {A.~G.~R.}\ \bibnamefont {Thomas}},\ and\
  \bibinfo {author} {\bibfnamefont {C.~P.}\ \bibnamefont {Ridgers}},\
  }\bibfield  {title} {\bibinfo {title} {Electron spin polarization in
  realistic trajectories around the magnetic node of two counter-propagating,
  circularly polarized, ultra-intense lasers},\ }\href
  {http://stacks.iop.org/0741-3335/60/i=6/a=064003} {\bibfield  {journal}
  {\bibinfo  {journal} {Plasma Phys. Control. Fusion}\ }\textbf {\bibinfo
  {volume} {60}},\ \bibinfo {pages} {064003} (\bibinfo {year}
  {2018})}\BibitemShut {NoStop}%
\bibitem [{\citenamefont {Li}\ \emph {et~al.}(2020)\citenamefont {Li},
  \citenamefont {Chen}, \citenamefont {Wang},\ and\ \citenamefont
  {Hu}}]{li2020production}%
  \BibitemOpen
  \bibfield  {author} {\bibinfo {author} {\bibfnamefont {Y.-F.}\ \bibnamefont
  {Li}}, \bibinfo {author} {\bibfnamefont {Y.-Y.}\ \bibnamefont {Chen}},
  \bibinfo {author} {\bibfnamefont {W.-M.}\ \bibnamefont {Wang}},\ and\
  \bibinfo {author} {\bibfnamefont {H.-S.}\ \bibnamefont {Hu}},\ }\bibfield
  {title} {\bibinfo {title} {Production of highly polarized positron beams via
  helicity transfer from polarized electrons in a strong laser field},\
  }\href@noop {} {\bibfield  {journal} {\bibinfo  {journal} {Physical Review
  Letters}\ }\textbf {\bibinfo {volume} {125}},\ \bibinfo {pages} {044802}
  (\bibinfo {year} {2020})}\BibitemShut {NoStop}%
\bibitem [{\citenamefont {Sokolov}\ and\ \citenamefont
  {Ternov}(1968)}]{Sokolov_1968}%
  \BibitemOpen
  \bibfield  {author} {\bibinfo {author} {\bibfnamefont {A.~A.}\ \bibnamefont
  {Sokolov}}\ and\ \bibinfo {author} {\bibfnamefont {I.~M.}\ \bibnamefont
  {Ternov}},\ }\href@noop {} {\emph {\bibinfo {title} {Synchrotron
  Radiation}}}\ (\bibinfo  {publisher} {Akademic, Germany},\ \bibinfo {year}
  {1968})\BibitemShut {NoStop}%
\bibitem [{\citenamefont {Baier}\ and\ \citenamefont
  {Katkov}(1967)}]{Baier_1967}%
  \BibitemOpen
  \bibfield  {author} {\bibinfo {author} {\bibfnamefont {V.~N.}\ \bibnamefont
  {Baier}}\ and\ \bibinfo {author} {\bibfnamefont {V.~M.}\ \bibnamefont
  {Katkov}},\ }\bibfield  {title} {\bibinfo {title} {{Radiational polarization
  of electrons in inhomogeneous magnetic field}},\ }\href@noop {} {\bibfield
  {journal} {\bibinfo  {journal} {Phys. Lett. A}\ }\textbf {\bibinfo {volume}
  {24}},\ \bibinfo {pages} {327} (\bibinfo {year} {1967})}\BibitemShut
  {NoStop}%
\bibitem [{\citenamefont {Baier}(1972)}]{Baier_1972}%
  \BibitemOpen
  \bibfield  {author} {\bibinfo {author} {\bibfnamefont {V.~N.}\ \bibnamefont
  {Baier}},\ }\bibfield  {title} {\bibinfo {title} {Radiative polarization of
  electron in storage rings},\ }\href@noop {} {\bibfield  {journal} {\bibinfo
  {journal} {Sov. Phys. Usp.}\ }\textbf {\bibinfo {volume} {14}},\ \bibinfo
  {pages} {695} (\bibinfo {year} {1972})}\BibitemShut {NoStop}%
\bibitem [{\citenamefont {Derbenev}\ and\ \citenamefont
  {Kondratenko}(1973)}]{Derbenev_1973}%
  \BibitemOpen
  \bibfield  {author} {\bibinfo {author} {\bibfnamefont {Y.}~\bibnamefont
  {Derbenev}}\ and\ \bibinfo {author} {\bibfnamefont {A.~M.}\ \bibnamefont
  {Kondratenko}},\ }\bibfield  {title} {\bibinfo {title} {{Polarization
  kinematics of particles in storage rings}},\ }\href@noop {} {\bibfield
  {journal} {\bibinfo  {journal} {Sov. Phys. JETP}\ }\textbf {\bibinfo {volume}
  {37}},\ \bibinfo {pages} {968} (\bibinfo {year} {1973})}\BibitemShut
  {NoStop}%
\bibitem [{\citenamefont {Li}\ \emph {et~al.}(2019{\natexlab{a}})\citenamefont
  {Li}, \citenamefont {Shaisultanov}, \citenamefont {Hatsagortsyan},
  \citenamefont {Wan}, \citenamefont {Keitel},\ and\ \citenamefont
  {Li}}]{li2019ultrarelativistic_PRL}%
  \BibitemOpen
  \bibfield  {author} {\bibinfo {author} {\bibfnamefont {Y.-F.}\ \bibnamefont
  {Li}}, \bibinfo {author} {\bibfnamefont {R.}~\bibnamefont {Shaisultanov}},
  \bibinfo {author} {\bibfnamefont {K.~Z.}\ \bibnamefont {Hatsagortsyan}},
  \bibinfo {author} {\bibfnamefont {F.}~\bibnamefont {Wan}}, \bibinfo {author}
  {\bibfnamefont {C.~H.}\ \bibnamefont {Keitel}},\ and\ \bibinfo {author}
  {\bibfnamefont {J.-X.}\ \bibnamefont {Li}},\ }\bibfield  {title} {\bibinfo
  {title} {Ultrarelativistic electron-beam polarization in single-shot
  interaction with an ultraintense laser pulse},\ }\href@noop {} {\bibfield
  {journal} {\bibinfo  {journal} {Physical review letters}\ }\textbf {\bibinfo
  {volume} {122}},\ \bibinfo {pages} {154801} (\bibinfo {year}
  {2019}{\natexlab{a}})}\BibitemShut {NoStop}%
\bibitem [{\citenamefont {Li}\ \emph {et~al.}(2019{\natexlab{b}})\citenamefont
  {Li}, \citenamefont {Guo}, \citenamefont {Shaisultanov}, \citenamefont
  {Hatsagortsyan},\ and\ \citenamefont {Li}}]{li2019electron_PRApp}%
  \BibitemOpen
  \bibfield  {author} {\bibinfo {author} {\bibfnamefont {Y.-F.}\ \bibnamefont
  {Li}}, \bibinfo {author} {\bibfnamefont {R.-T.}\ \bibnamefont {Guo}},
  \bibinfo {author} {\bibfnamefont {R.}~\bibnamefont {Shaisultanov}}, \bibinfo
  {author} {\bibfnamefont {K.~Z.}\ \bibnamefont {Hatsagortsyan}},\ and\
  \bibinfo {author} {\bibfnamefont {J.-X.}\ \bibnamefont {Li}},\ }\bibfield
  {title} {\bibinfo {title} {Electron polarimetry with nonlinear compton
  scattering},\ }\href@noop {} {\bibfield  {journal} {\bibinfo  {journal}
  {Physical Review Applied}\ }\textbf {\bibinfo {volume} {12}},\ \bibinfo
  {pages} {014047} (\bibinfo {year} {2019}{\natexlab{b}})}\BibitemShut
  {NoStop}%
\bibitem [{\citenamefont {Wan}\ \emph {et~al.}(2020)\citenamefont {Wan},
  \citenamefont {Shaisultanov}, \citenamefont {Li}, \citenamefont
  {Hatsagortsyan}, \citenamefont {Keitel},\ and\ \citenamefont
  {Li}}]{Wan_2019plb}%
  \BibitemOpen
  \bibfield  {author} {\bibinfo {author} {\bibfnamefont {F.}~\bibnamefont
  {Wan}}, \bibinfo {author} {\bibfnamefont {R.}~\bibnamefont {Shaisultanov}},
  \bibinfo {author} {\bibfnamefont {Y.-F.}\ \bibnamefont {Li}}, \bibinfo
  {author} {\bibfnamefont {K.~Z.}\ \bibnamefont {Hatsagortsyan}}, \bibinfo
  {author} {\bibfnamefont {C.~H.}\ \bibnamefont {Keitel}},\ and\ \bibinfo
  {author} {\bibfnamefont {J.-X.}\ \bibnamefont {Li}},\ }\bibfield  {title}
  {\bibinfo {title} {Ultrarelativistic polarized positron jets via collision of
  electron and ultraintense laser beams},\ }\href@noop {} {\bibfield  {journal}
  {\bibinfo  {journal} {Phys. Lett. B}\ }\textbf {\bibinfo {volume} {800}},\
  \bibinfo {pages} {135120} (\bibinfo {year} {2020})}\BibitemShut {NoStop}%
\bibitem [{\citenamefont {Chen}\ \emph {et~al.}(2019)\citenamefont {Chen},
  \citenamefont {He}, \citenamefont {Shaisultanov}, \citenamefont
  {Hatsagortsyan},\ and\ \citenamefont {Keitel}}]{chen2019polarized}%
  \BibitemOpen
  \bibfield  {author} {\bibinfo {author} {\bibfnamefont {Y.-Y.}\ \bibnamefont
  {Chen}}, \bibinfo {author} {\bibfnamefont {P.-L.}\ \bibnamefont {He}},
  \bibinfo {author} {\bibfnamefont {R.}~\bibnamefont {Shaisultanov}}, \bibinfo
  {author} {\bibfnamefont {K.~Z.}\ \bibnamefont {Hatsagortsyan}},\ and\
  \bibinfo {author} {\bibfnamefont {C.~H.}\ \bibnamefont {Keitel}},\ }\bibfield
   {title} {\bibinfo {title} {Polarized positron beams via intense two-color
  laser pulses},\ }\href@noop {} {\bibfield  {journal} {\bibinfo  {journal}
  {Physical review letters}\ }\textbf {\bibinfo {volume} {123}},\ \bibinfo
  {pages} {174801} (\bibinfo {year} {2019})}\BibitemShut {NoStop}%
\bibitem [{\citenamefont {Seipt}\ \emph {et~al.}(2019)\citenamefont {Seipt},
  \citenamefont {Del~Sorbo}, \citenamefont {Ridgers},\ and\ \citenamefont
  {Thomas}}]{Seipt_2019}%
  \BibitemOpen
  \bibfield  {author} {\bibinfo {author} {\bibfnamefont {D.}~\bibnamefont
  {Seipt}}, \bibinfo {author} {\bibfnamefont {D.}~\bibnamefont {Del~Sorbo}},
  \bibinfo {author} {\bibfnamefont {C.~P.}\ \bibnamefont {Ridgers}},\ and\
  \bibinfo {author} {\bibfnamefont {A.~G.~R.}\ \bibnamefont {Thomas}},\
  }\bibfield  {title} {\bibinfo {title} {Ultrafast polarization of an electron
  beam in an intense bichromatic laser field},\ }\href
  {https://doi.org/10.1103/PhysRevA.100.061402} {\bibfield  {journal} {\bibinfo
   {journal} {Phys. Rev. A}\ }\textbf {\bibinfo {volume} {100}},\ \bibinfo
  {pages} {061402(R)} (\bibinfo {year} {2019})}\BibitemShut {NoStop}%
\bibitem [{\citenamefont {Pukhov}\ and\ \citenamefont {Meyer-ter
  Vehn}(1996)}]{pukhov1996relativistic}%
  \BibitemOpen
  \bibfield  {author} {\bibinfo {author} {\bibfnamefont {A.}~\bibnamefont
  {Pukhov}}\ and\ \bibinfo {author} {\bibfnamefont {J.}~\bibnamefont {Meyer-ter
  Vehn}},\ }\bibfield  {title} {\bibinfo {title} {Relativistic magnetic
  self-channeling of light in near-critical plasma: three-dimensional
  particle-in-cell simulation},\ }\href@noop {} {\bibfield  {journal} {\bibinfo
   {journal} {Physical review letters}\ }\textbf {\bibinfo {volume} {76}},\
  \bibinfo {pages} {3975} (\bibinfo {year} {1996})}\BibitemShut {NoStop}%
\bibitem [{\citenamefont {Pukhov}\ \emph {et~al.}(1999)\citenamefont {Pukhov},
  \citenamefont {Sheng},\ and\ \citenamefont {Meyer-ter
  Vehn}}]{pukhov1999_DLA}%
  \BibitemOpen
  \bibfield  {author} {\bibinfo {author} {\bibfnamefont {A.}~\bibnamefont
  {Pukhov}}, \bibinfo {author} {\bibfnamefont {Z.-M.}\ \bibnamefont {Sheng}},\
  and\ \bibinfo {author} {\bibfnamefont {J.}~\bibnamefont {Meyer-ter Vehn}},\
  }\bibfield  {title} {\bibinfo {title} {Particle acceleration in relativistic
  laser channels},\ }\href@noop {} {\bibfield  {journal} {\bibinfo  {journal}
  {Physics of Plasmas}\ }\textbf {\bibinfo {volume} {6}},\ \bibinfo {pages}
  {2847} (\bibinfo {year} {1999})}\BibitemShut {NoStop}%
\bibitem [{\citenamefont {Stark}\ \emph {et~al.}(2016)\citenamefont {Stark},
  \citenamefont {Toncian},\ and\ \citenamefont {Arefiev}}]{Stark2016_PRL}%
  \BibitemOpen
  \bibfield  {author} {\bibinfo {author} {\bibfnamefont {D.}~\bibnamefont
  {Stark}}, \bibinfo {author} {\bibfnamefont {T.}~\bibnamefont {Toncian}},\
  and\ \bibinfo {author} {\bibfnamefont {A.}~\bibnamefont {Arefiev}},\
  }\bibfield  {title} {\bibinfo {title} {Enhanced multi-mev photon emission by
  a laser-driven electron beam in a self-generated magnetic field},\
  }\href@noop {} {\bibfield  {journal} {\bibinfo  {journal} {Physical review
  letters}\ }\textbf {\bibinfo {volume} {116}},\ \bibinfo {pages} {185003}
  (\bibinfo {year} {2016})}\BibitemShut {NoStop}%
\bibitem [{\citenamefont {Gong}\ \emph {et~al.}(2020)\citenamefont {Gong},
  \citenamefont {Mackenroth}, \citenamefont {Wang}, \citenamefont {Yan},
  \citenamefont {Toncian},\ and\ \citenamefont {Arefiev}}]{gong2020direct}%
  \BibitemOpen
  \bibfield  {author} {\bibinfo {author} {\bibfnamefont {Z.}~\bibnamefont
  {Gong}}, \bibinfo {author} {\bibfnamefont {F.}~\bibnamefont {Mackenroth}},
  \bibinfo {author} {\bibfnamefont {T.}~\bibnamefont {Wang}}, \bibinfo {author}
  {\bibfnamefont {X.}~\bibnamefont {Yan}}, \bibinfo {author} {\bibfnamefont
  {T.}~\bibnamefont {Toncian}},\ and\ \bibinfo {author} {\bibfnamefont
  {A.}~\bibnamefont {Arefiev}},\ }\bibfield  {title} {\bibinfo {title} {Direct
  laser acceleration of electrons assisted by strong laser-driven azimuthal
  plasma magnetic fields},\ }\href@noop {} {\bibfield  {journal} {\bibinfo
  {journal} {Physical Review E}\ }\textbf {\bibinfo {volume} {102}},\ \bibinfo
  {pages} {013206} (\bibinfo {year} {2020})}\BibitemShut {NoStop}%
\bibitem [{\citenamefont {Hussein}\ \emph {et~al.}(2021)\citenamefont
  {Hussein}, \citenamefont {Arefiev}, \citenamefont {Batson}, \citenamefont
  {Chen}, \citenamefont {Craxton} \emph {et~al.}}]{hussein2021towards}%
  \BibitemOpen
  \bibfield  {author} {\bibinfo {author} {\bibfnamefont {A.~E.}\ \bibnamefont
  {Hussein}}, \bibinfo {author} {\bibfnamefont {A.~V.}\ \bibnamefont
  {Arefiev}}, \bibinfo {author} {\bibfnamefont {T.}~\bibnamefont {Batson}},
  \bibinfo {author} {\bibfnamefont {H.}~\bibnamefont {Chen}}, \bibinfo {author}
  {\bibfnamefont {R.}~\bibnamefont {Craxton}}, \emph {et~al.},\ }\bibfield
  {title} {\bibinfo {title} {Towards the optimisation of direct laser
  acceleration},\ }\href@noop {} {\bibfield  {journal} {\bibinfo  {journal}
  {New Journal of Physics}\ }\textbf {\bibinfo {volume} {23}},\ \bibinfo
  {pages} {023031} (\bibinfo {year} {2021})}\BibitemShut {NoStop}%
\bibitem [{\citenamefont {Arber}\ \emph {et~al.}(2015)\citenamefont {Arber},
  \citenamefont {Bennett}, \citenamefont {Brady}, \citenamefont
  {Lawrence-Douglas}, \citenamefont {Ramsay}, \citenamefont {Sircombe},
  \citenamefont {Gillies}, \citenamefont {Evans}, \citenamefont {Schmitz},
  \citenamefont {Bell} \emph {et~al.}}]{arber2015contemporary}%
  \BibitemOpen
  \bibfield  {author} {\bibinfo {author} {\bibfnamefont {T.}~\bibnamefont
  {Arber}}, \bibinfo {author} {\bibfnamefont {K.}~\bibnamefont {Bennett}},
  \bibinfo {author} {\bibfnamefont {C.}~\bibnamefont {Brady}}, \bibinfo
  {author} {\bibfnamefont {A.}~\bibnamefont {Lawrence-Douglas}}, \bibinfo
  {author} {\bibfnamefont {M.}~\bibnamefont {Ramsay}}, \bibinfo {author}
  {\bibfnamefont {N.}~\bibnamefont {Sircombe}}, \bibinfo {author}
  {\bibfnamefont {P.}~\bibnamefont {Gillies}}, \bibinfo {author} {\bibfnamefont
  {R.}~\bibnamefont {Evans}}, \bibinfo {author} {\bibfnamefont
  {H.}~\bibnamefont {Schmitz}}, \bibinfo {author} {\bibfnamefont
  {A.}~\bibnamefont {Bell}}, \emph {et~al.},\ }\bibfield  {title} {\bibinfo
  {title} {Contemporary particle-in-cell approach to laser-plasma modelling},\
  }\href@noop {} {\bibfield  {journal} {\bibinfo  {journal} {Plasma Physics and
  Controlled Fusion}\ }\textbf {\bibinfo {volume} {57}},\ \bibinfo {pages}
  {113001} (\bibinfo {year} {2015})}\BibitemShut {NoStop}%
\bibitem [{SM()}]{SM}%
  \BibitemOpen
  \href@noop {} {}\bibinfo {howpublished} {See the Supplemental Materials for
  the details.}\BibitemShut {Stop}%
\bibitem [{\citenamefont {Pukhov}(2002)}]{pukhov2002strong}%
  \BibitemOpen
  \bibfield  {author} {\bibinfo {author} {\bibfnamefont {A.}~\bibnamefont
  {Pukhov}},\ }\bibfield  {title} {\bibinfo {title} {Strong field interaction
  of laser radiation},\ }\href@noop {} {\bibfield  {journal} {\bibinfo
  {journal} {Reports on progress in Physics}\ }\textbf {\bibinfo {volume}
  {66}},\ \bibinfo {pages} {47} (\bibinfo {year} {2002})}\BibitemShut {NoStop}%
\bibitem [{\citenamefont {Ji}\ \emph {et~al.}(2014)\citenamefont {Ji},
  \citenamefont {Pukhov}, \citenamefont {Kostyukov}, \citenamefont {Shen},\
  and\ \citenamefont {Akli}}]{ji2014_PRL}%
  \BibitemOpen
  \bibfield  {author} {\bibinfo {author} {\bibfnamefont {L.}~\bibnamefont
  {Ji}}, \bibinfo {author} {\bibfnamefont {A.}~\bibnamefont {Pukhov}}, \bibinfo
  {author} {\bibfnamefont {I.~Y.}\ \bibnamefont {Kostyukov}}, \bibinfo {author}
  {\bibfnamefont {B.}~\bibnamefont {Shen}},\ and\ \bibinfo {author}
  {\bibfnamefont {K.}~\bibnamefont {Akli}},\ }\bibfield  {title} {\bibinfo
  {title} {Radiation-reaction trapping of electrons in extreme laser fields},\
  }\href@noop {} {\bibfield  {journal} {\bibinfo  {journal} {Physical review
  letters}\ }\textbf {\bibinfo {volume} {112}},\ \bibinfo {pages} {145003}
  (\bibinfo {year} {2014})}\BibitemShut {NoStop}%
\bibitem [{\citenamefont {Jansen}\ \emph {et~al.}(2018)\citenamefont {Jansen},
  \citenamefont {Wang}, \citenamefont {Stark}, \citenamefont
  {d’Humi{\`e}res}, \citenamefont {Toncian},\ and\ \citenamefont
  {Arefiev}}]{jansen2018leveraging}%
  \BibitemOpen
  \bibfield  {author} {\bibinfo {author} {\bibfnamefont {O.}~\bibnamefont
  {Jansen}}, \bibinfo {author} {\bibfnamefont {T.}~\bibnamefont {Wang}},
  \bibinfo {author} {\bibfnamefont {D.~J.}\ \bibnamefont {Stark}}, \bibinfo
  {author} {\bibfnamefont {E.}~\bibnamefont {d’Humi{\`e}res}}, \bibinfo
  {author} {\bibfnamefont {T.}~\bibnamefont {Toncian}},\ and\ \bibinfo {author}
  {\bibfnamefont {A.}~\bibnamefont {Arefiev}},\ }\bibfield  {title} {\bibinfo
  {title} {Leveraging extreme laser-driven magnetic fields for gamma-ray
  generation and pair production},\ }\href@noop {} {\bibfield  {journal}
  {\bibinfo  {journal} {Plasma Physics and Controlled Fusion}\ }\textbf
  {\bibinfo {volume} {60}},\ \bibinfo {pages} {054006} (\bibinfo {year}
  {2018})}\BibitemShut {NoStop}%
\bibitem [{\citenamefont {Narayan}\ \emph {et~al.}(2016)\citenamefont
  {Narayan}, \citenamefont {Jones}, \citenamefont {Cornejo}, \citenamefont
  {Dalton}, \citenamefont {Deconinck}, \citenamefont {Dutta}, \citenamefont
  {Gaskell}, \citenamefont {Martin}, \citenamefont {Paschke}, \citenamefont
  {Tvaskis} \emph {et~al.}}]{narayan2016precision}%
  \BibitemOpen
  \bibfield  {author} {\bibinfo {author} {\bibfnamefont {A.}~\bibnamefont
  {Narayan}}, \bibinfo {author} {\bibfnamefont {D.}~\bibnamefont {Jones}},
  \bibinfo {author} {\bibfnamefont {J.}~\bibnamefont {Cornejo}}, \bibinfo
  {author} {\bibfnamefont {M.}~\bibnamefont {Dalton}}, \bibinfo {author}
  {\bibfnamefont {W.}~\bibnamefont {Deconinck}}, \bibinfo {author}
  {\bibfnamefont {D.}~\bibnamefont {Dutta}}, \bibinfo {author} {\bibfnamefont
  {D.}~\bibnamefont {Gaskell}}, \bibinfo {author} {\bibfnamefont
  {J.}~\bibnamefont {Martin}}, \bibinfo {author} {\bibfnamefont
  {K.}~\bibnamefont {Paschke}}, \bibinfo {author} {\bibfnamefont
  {V.}~\bibnamefont {Tvaskis}}, \emph {et~al.},\ }\bibfield  {title} {\bibinfo
  {title} {Precision electron-beam polarimetry at 1 gev using diamond
  microstrip detectors},\ }\href@noop {} {\bibfield  {journal} {\bibinfo
  {journal} {Physical Review X}\ }\textbf {\bibinfo {volume} {6}},\ \bibinfo
  {pages} {011013} (\bibinfo {year} {2016})}\BibitemShut {NoStop}%
\bibitem [{\citenamefont {Zhang}\ \emph
  {et~al.}(2020{\natexlab{b}})\citenamefont {Zhang}, \citenamefont {Bulanov},
  \citenamefont {Seipt}, \citenamefont {Arefiev},\ and\ \citenamefont
  {Thomas}}]{zhang2020relativistic}%
  \BibitemOpen
  \bibfield  {author} {\bibinfo {author} {\bibfnamefont {P.}~\bibnamefont
  {Zhang}}, \bibinfo {author} {\bibfnamefont {S.}~\bibnamefont {Bulanov}},
  \bibinfo {author} {\bibfnamefont {D.}~\bibnamefont {Seipt}}, \bibinfo
  {author} {\bibfnamefont {A.}~\bibnamefont {Arefiev}},\ and\ \bibinfo {author}
  {\bibfnamefont {A.}~\bibnamefont {Thomas}},\ }\bibfield  {title} {\bibinfo
  {title} {Relativistic plasma physics in supercritical fields},\ }\href@noop
  {} {\bibfield  {journal} {\bibinfo  {journal} {Physics of Plasmas}\ }\textbf
  {\bibinfo {volume} {27}},\ \bibinfo {pages} {050601} (\bibinfo {year}
  {2020}{\natexlab{b}})}\BibitemShut {NoStop}%
\bibitem [{\citenamefont {Thomas}(1927)}]{thomas1927kinematics}%
  \BibitemOpen
  \bibfield  {author} {\bibinfo {author} {\bibfnamefont {L.~H.}\ \bibnamefont
  {Thomas}},\ }\bibfield  {title} {\bibinfo {title} {I. the kinematics of an
  electron with an axis},\ }\href@noop {} {\bibfield  {journal} {\bibinfo
  {journal} {The London, Edinburgh, and Dublin Philosophical Magazine and
  Journal of Science}\ }\textbf {\bibinfo {volume} {3}},\ \bibinfo {pages} {1}
  (\bibinfo {year} {1927})}\BibitemShut {NoStop}%
\bibitem [{\citenamefont {Bargmann}\ \emph {et~al.}(1959)\citenamefont
  {Bargmann}, \citenamefont {Michel},\ and\ \citenamefont
  {Telegdi}}]{bargmann1959precession}%
  \BibitemOpen
  \bibfield  {author} {\bibinfo {author} {\bibfnamefont {V.}~\bibnamefont
  {Bargmann}}, \bibinfo {author} {\bibfnamefont {L.}~\bibnamefont {Michel}},\
  and\ \bibinfo {author} {\bibfnamefont {V.}~\bibnamefont {Telegdi}},\
  }\bibfield  {title} {\bibinfo {title} {Precession of the polarization of
  particles moving in a homogeneous electromagnetic field},\ }\href@noop {}
  {\bibfield  {journal} {\bibinfo  {journal} {Physical Review Letters}\
  }\textbf {\bibinfo {volume} {2}},\ \bibinfo {pages} {435} (\bibinfo {year}
  {1959})}\BibitemShut {NoStop}%
\bibitem [{\citenamefont {Duclous}\ \emph {et~al.}(2010)\citenamefont
  {Duclous}, \citenamefont {Kirk},\ and\ \citenamefont
  {Bell}}]{duclous2010monte}%
  \BibitemOpen
  \bibfield  {author} {\bibinfo {author} {\bibfnamefont {R.}~\bibnamefont
  {Duclous}}, \bibinfo {author} {\bibfnamefont {J.~G.}\ \bibnamefont {Kirk}},\
  and\ \bibinfo {author} {\bibfnamefont {A.~R.}\ \bibnamefont {Bell}},\
  }\bibfield  {title} {\bibinfo {title} {Monte carlo calculations of pair
  production in high-intensity laser--plasma interactions},\ }\href@noop {}
  {\bibfield  {journal} {\bibinfo  {journal} {Plasma Physics and Controlled
  Fusion}\ }\textbf {\bibinfo {volume} {53}},\ \bibinfo {pages} {015009}
  (\bibinfo {year} {2010})}\BibitemShut {NoStop}%
\bibitem [{\citenamefont {Elkina}\ \emph {et~al.}(2011)\citenamefont {Elkina},
  \citenamefont {Fedotov}, \citenamefont {Kostyukov}, \citenamefont {Legkov},
  \citenamefont {Narozhny}, \citenamefont {Nerush},\ and\ \citenamefont
  {Ruhl}}]{elkina2011qed}%
  \BibitemOpen
  \bibfield  {author} {\bibinfo {author} {\bibfnamefont {N.}~\bibnamefont
  {Elkina}}, \bibinfo {author} {\bibfnamefont {A.}~\bibnamefont {Fedotov}},
  \bibinfo {author} {\bibfnamefont {I.~Y.}\ \bibnamefont {Kostyukov}}, \bibinfo
  {author} {\bibfnamefont {M.}~\bibnamefont {Legkov}}, \bibinfo {author}
  {\bibfnamefont {N.}~\bibnamefont {Narozhny}}, \bibinfo {author}
  {\bibfnamefont {E.}~\bibnamefont {Nerush}},\ and\ \bibinfo {author}
  {\bibfnamefont {H.}~\bibnamefont {Ruhl}},\ }\bibfield  {title} {\bibinfo
  {title} {Qed cascades induced by circularly polarized laser fields},\
  }\href@noop {} {\bibfield  {journal} {\bibinfo  {journal} {Physical Review
  Special Topics-Accelerators and Beams}\ }\textbf {\bibinfo {volume} {14}},\
  \bibinfo {pages} {054401} (\bibinfo {year} {2011})}\BibitemShut {NoStop}%
\bibitem [{\citenamefont {Ridgers}\ \emph {et~al.}(2014)\citenamefont
  {Ridgers}, \citenamefont {Kirk}, \citenamefont {Duclous}, \citenamefont
  {Blackburn}, \citenamefont {Brady}, \citenamefont {Bennett}, \citenamefont
  {Arber},\ and\ \citenamefont {Bell}}]{ridgers2014modelling}%
  \BibitemOpen
  \bibfield  {author} {\bibinfo {author} {\bibfnamefont {C.}~\bibnamefont
  {Ridgers}}, \bibinfo {author} {\bibfnamefont {J.~G.}\ \bibnamefont {Kirk}},
  \bibinfo {author} {\bibfnamefont {R.}~\bibnamefont {Duclous}}, \bibinfo
  {author} {\bibfnamefont {T.}~\bibnamefont {Blackburn}}, \bibinfo {author}
  {\bibfnamefont {C.}~\bibnamefont {Brady}}, \bibinfo {author} {\bibfnamefont
  {K.}~\bibnamefont {Bennett}}, \bibinfo {author} {\bibfnamefont
  {T.}~\bibnamefont {Arber}},\ and\ \bibinfo {author} {\bibfnamefont
  {A.}~\bibnamefont {Bell}},\ }\bibfield  {title} {\bibinfo {title} {Modelling
  gamma-ray photon emission and pair production in high-intensity laser--matter
  interactions},\ }\href@noop {} {\bibfield  {journal} {\bibinfo  {journal}
  {Journal of Computational Physics}\ }\textbf {\bibinfo {volume} {260}},\
  \bibinfo {pages} {273} (\bibinfo {year} {2014})}\BibitemShut {NoStop}%
\bibitem [{\citenamefont {Gonoskov}\ \emph {et~al.}(2015)\citenamefont
  {Gonoskov}, \citenamefont {Bastrakov}, \citenamefont {Efimenko},
  \citenamefont {Ilderton}, \citenamefont {Marklund}, \citenamefont {Meyerov},
  \citenamefont {Muraviev}, \citenamefont {Sergeev}, \citenamefont {Surmin},\
  and\ \citenamefont {Wallin}}]{gonoskov2015extended}%
  \BibitemOpen
  \bibfield  {author} {\bibinfo {author} {\bibfnamefont {A.}~\bibnamefont
  {Gonoskov}}, \bibinfo {author} {\bibfnamefont {S.}~\bibnamefont {Bastrakov}},
  \bibinfo {author} {\bibfnamefont {E.}~\bibnamefont {Efimenko}}, \bibinfo
  {author} {\bibfnamefont {A.}~\bibnamefont {Ilderton}}, \bibinfo {author}
  {\bibfnamefont {M.}~\bibnamefont {Marklund}}, \bibinfo {author}
  {\bibfnamefont {I.}~\bibnamefont {Meyerov}}, \bibinfo {author} {\bibfnamefont
  {A.}~\bibnamefont {Muraviev}}, \bibinfo {author} {\bibfnamefont
  {A.}~\bibnamefont {Sergeev}}, \bibinfo {author} {\bibfnamefont
  {I.}~\bibnamefont {Surmin}},\ and\ \bibinfo {author} {\bibfnamefont
  {E.}~\bibnamefont {Wallin}},\ }\bibfield  {title} {\bibinfo {title} {Extended
  particle-in-cell schemes for physics in ultrastrong laser fields: Review and
  developments},\ }\href@noop {} {\bibfield  {journal} {\bibinfo  {journal}
  {Physical Review E}\ }\textbf {\bibinfo {volume} {92}},\ \bibinfo {pages}
  {023305} (\bibinfo {year} {2015})}\BibitemShut {NoStop}%
\bibitem [{\citenamefont {Ma}\ \emph {et~al.}(2007)\citenamefont {Ma},
  \citenamefont {Song}, \citenamefont {Yang}, \citenamefont {Zhang},
  \citenamefont {Zhao}, \citenamefont {Sun}, \citenamefont {Ren}, \citenamefont
  {Liu}, \citenamefont {Liu}, \citenamefont {Shen} \emph
  {et~al.}}]{ma2007directly}%
  \BibitemOpen
  \bibfield  {author} {\bibinfo {author} {\bibfnamefont {W.}~\bibnamefont
  {Ma}}, \bibinfo {author} {\bibfnamefont {L.}~\bibnamefont {Song}}, \bibinfo
  {author} {\bibfnamefont {R.}~\bibnamefont {Yang}}, \bibinfo {author}
  {\bibfnamefont {T.}~\bibnamefont {Zhang}}, \bibinfo {author} {\bibfnamefont
  {Y.}~\bibnamefont {Zhao}}, \bibinfo {author} {\bibfnamefont {L.}~\bibnamefont
  {Sun}}, \bibinfo {author} {\bibfnamefont {Y.}~\bibnamefont {Ren}}, \bibinfo
  {author} {\bibfnamefont {D.}~\bibnamefont {Liu}}, \bibinfo {author}
  {\bibfnamefont {L.}~\bibnamefont {Liu}}, \bibinfo {author} {\bibfnamefont
  {J.}~\bibnamefont {Shen}}, \emph {et~al.},\ }\bibfield  {title} {\bibinfo
  {title} {Directly synthesized strong, highly conducting, transparent
  single-walled carbon nanotube films},\ }\href@noop {} {\bibfield  {journal}
  {\bibinfo  {journal} {Nano Letters}\ }\textbf {\bibinfo {volume} {7}},\
  \bibinfo {pages} {2307} (\bibinfo {year} {2007})}\BibitemShut {NoStop}%
\bibitem [{\citenamefont {Bin}\ \emph {et~al.}(2015)\citenamefont {Bin},
  \citenamefont {Ma}, \citenamefont {Wang}, \citenamefont {Streeter},
  \citenamefont {Kreuzer}, \citenamefont {Kiefer}, \citenamefont {Yeung},
  \citenamefont {Cousens}, \citenamefont {Foster}, \citenamefont {Dromey} \emph
  {et~al.}}]{bin2015ion}%
  \BibitemOpen
  \bibfield  {author} {\bibinfo {author} {\bibfnamefont {J.}~\bibnamefont
  {Bin}}, \bibinfo {author} {\bibfnamefont {W.}~\bibnamefont {Ma}}, \bibinfo
  {author} {\bibfnamefont {H.}~\bibnamefont {Wang}}, \bibinfo {author}
  {\bibfnamefont {M.}~\bibnamefont {Streeter}}, \bibinfo {author}
  {\bibfnamefont {C.}~\bibnamefont {Kreuzer}}, \bibinfo {author} {\bibfnamefont
  {D.}~\bibnamefont {Kiefer}}, \bibinfo {author} {\bibfnamefont
  {M.}~\bibnamefont {Yeung}}, \bibinfo {author} {\bibfnamefont
  {S.}~\bibnamefont {Cousens}}, \bibinfo {author} {\bibfnamefont
  {P.}~\bibnamefont {Foster}}, \bibinfo {author} {\bibfnamefont
  {B.}~\bibnamefont {Dromey}}, \emph {et~al.},\ }\bibfield  {title} {\bibinfo
  {title} {Ion acceleration using relativistic pulse shaping in
  near-critical-density plasmas},\ }\href@noop {} {\bibfield  {journal}
  {\bibinfo  {journal} {Physical review letters}\ }\textbf {\bibinfo {volume}
  {115}},\ \bibinfo {pages} {064801} (\bibinfo {year} {2015})}\BibitemShut
  {NoStop}%
\bibitem [{\citenamefont {Wang}\ \emph {et~al.}(2011)\citenamefont {Wang},
  \citenamefont {Lin}, \citenamefont {Sheng}, \citenamefont {Liu},
  \citenamefont {Zhao}, \citenamefont {Guo}, \citenamefont {Lu}, \citenamefont
  {He}, \citenamefont {Chen},\ and\ \citenamefont {Yan}}]{wang2011laser}%
  \BibitemOpen
  \bibfield  {author} {\bibinfo {author} {\bibfnamefont {H.}~\bibnamefont
  {Wang}}, \bibinfo {author} {\bibfnamefont {C.}~\bibnamefont {Lin}}, \bibinfo
  {author} {\bibfnamefont {Z.}~\bibnamefont {Sheng}}, \bibinfo {author}
  {\bibfnamefont {B.}~\bibnamefont {Liu}}, \bibinfo {author} {\bibfnamefont
  {S.}~\bibnamefont {Zhao}}, \bibinfo {author} {\bibfnamefont {Z.}~\bibnamefont
  {Guo}}, \bibinfo {author} {\bibfnamefont {Y.}~\bibnamefont {Lu}}, \bibinfo
  {author} {\bibfnamefont {X.}~\bibnamefont {He}}, \bibinfo {author}
  {\bibfnamefont {J.}~\bibnamefont {Chen}},\ and\ \bibinfo {author}
  {\bibfnamefont {X.}~\bibnamefont {Yan}},\ }\bibfield  {title} {\bibinfo
  {title} {Laser shaping of a relativistic intense, short gaussian pulse by a
  plasma lens},\ }\href@noop {} {\bibfield  {journal} {\bibinfo  {journal}
  {Physical review letters}\ }\textbf {\bibinfo {volume} {107}},\ \bibinfo
  {pages} {265002} (\bibinfo {year} {2011})}\BibitemShut {NoStop}%
\end{thebibliography}
